\documentclass[sigconf]{acmart}

\usepackage{caption}
\usepackage{subcaption}
\usepackage{tabularx}
\usepackage{makecell}
\usepackage{hyperref}
\usepackage{csquotes}

\usepackage{xcolor}

\copyrightyear{2026}
\acmYear{2026}
\setcopyright{cc}
\setcctype{by-nc-nd}
\acmConference[CHI '26]{Proceedings of the 2026 CHI Conference on Human Factors in Computing Systems}{April 13--17, 2026}{Barcelona, Spain}
\acmBooktitle{Proceedings of the 2026 CHI Conference on Human Factors in Computing Systems (CHI '26), April 13--17, 2026, Barcelona, Spain}
\acmPrice{}
\acmDOI{10.1145/3772318.3790651}
\acmISBN{979-8-4007-2278-3/2026/04}

\begin{document}




\title{Nudging the Somas: Exploring How Live-Configurable Mixed Reality Objects Shape Open-Ended Intercorporeal Movements}


\author{Botao Amber Hu}\authornote{Corresponding author}
\orcid{0000-0002-4504-0941}
\affiliation{
  \institution{Reality Design Lab}
  \city{New York City}
  \country{USA}
}
\affiliation{
  \institution{University of Oxford}
  \city{Oxford}
  \country{UK}
}
\email{botao@reality.design}

\author{Yilan Elan Tao}
\orcid{0000-0003-1691-9727}
\affiliation{%
 \institution{Simon Fraser University}
 \city{Vancouver}
 \country{Canada}
}
\email{elan_tao@sfu.ca}

\author{Rem RunGu Lin}
\orcid{0000-0003-1931-7609}
\affiliation{%
  \institution{The Hong Kong University of Science and Technology (Guangzhou)}
  \city{Guangzhou}
  \country{China}
  }
\email{rlin408@connect.hkust-gz.edu.cn}

\author{Mingze Chai}
\orcid{0009-0008-7168-8929}
\affiliation{%
 \institution{Independent}
 \city{Shanghai}
 \country{China}
}
\email{chaimignze@gmail.com}

\author{Yuemin Huang}
\orcid{0009-0009-5559-6474}
\affiliation{%
 \institution{East China Normal University}
 \city{Shanghai}
 \country{China}
}
\email{uminh1024@gmail.com}

\author{Rakesh Patibanda}
\orcid{0000-0002-2501-9969}
\affiliation{%
  \institution{Monash University}
  \city{Melbourne}
  \country{Australia}
}
\email{rakesh@exertiongameslab.org}



\begin{abstract}



Mixed Reality (MR) increasingly explores how virtual elements can shape physical behavior, yet how MR objects guide group movement remains underexplored. We address this gap by examining how virtual objects can nudge collective, co-located movement without relying on explicit instructions or choreography. We developed GravField, a research-through-design, co-located MR performance system where an “object jockey” live-configures virtual objects (e.g., ropes, springs, magnetic fields) with real-time, parameterized “digital physics” (e.g., weight, elasticity, force) to influence headset-wearing participants' movement, made perceptible through augmented visual and audio feedback serving as cognitive-somatic cues. Our bricolage analysis of the performances, based on video, interviews, soma trajectories, and field notes, indicates that these live nudges support emergent intercorporeal coordination and that ambiguity and real-time configuration sustain open-ended, exploratory engagement. Ultimately, our work offers empirical insights and design principles for MR systems that can guide group movement through embodied, felt dynamics while preserving participants’ sense of agency.
\end{abstract}

\ccsdesc[500]{Human-centered computing~Mixed / augmented reality}
\ccsdesc[500]{Human-centered computing~Collaborative interaction}
\ccsdesc[300]{Human-centered computing~Interaction design theory, concepts and paradigms}
\ccsdesc[300]{Human-centered computing~Empirical studies in HCI}
\ccsdesc[100]{Human-centered computing~Sound-based input / output}
\ccsdesc[100]{Applied computing~Performing arts}

\keywords{Nudge, Co-located Mixed Reality, Soma Design, Intercorporeal Design, Digital Object, Digital Physics, Live-Coding System, Performance, Open-endedness, Improvisation, Movement Computing}


\begin{teaserfigure}
    \centering
    \includegraphics[width=1\linewidth]{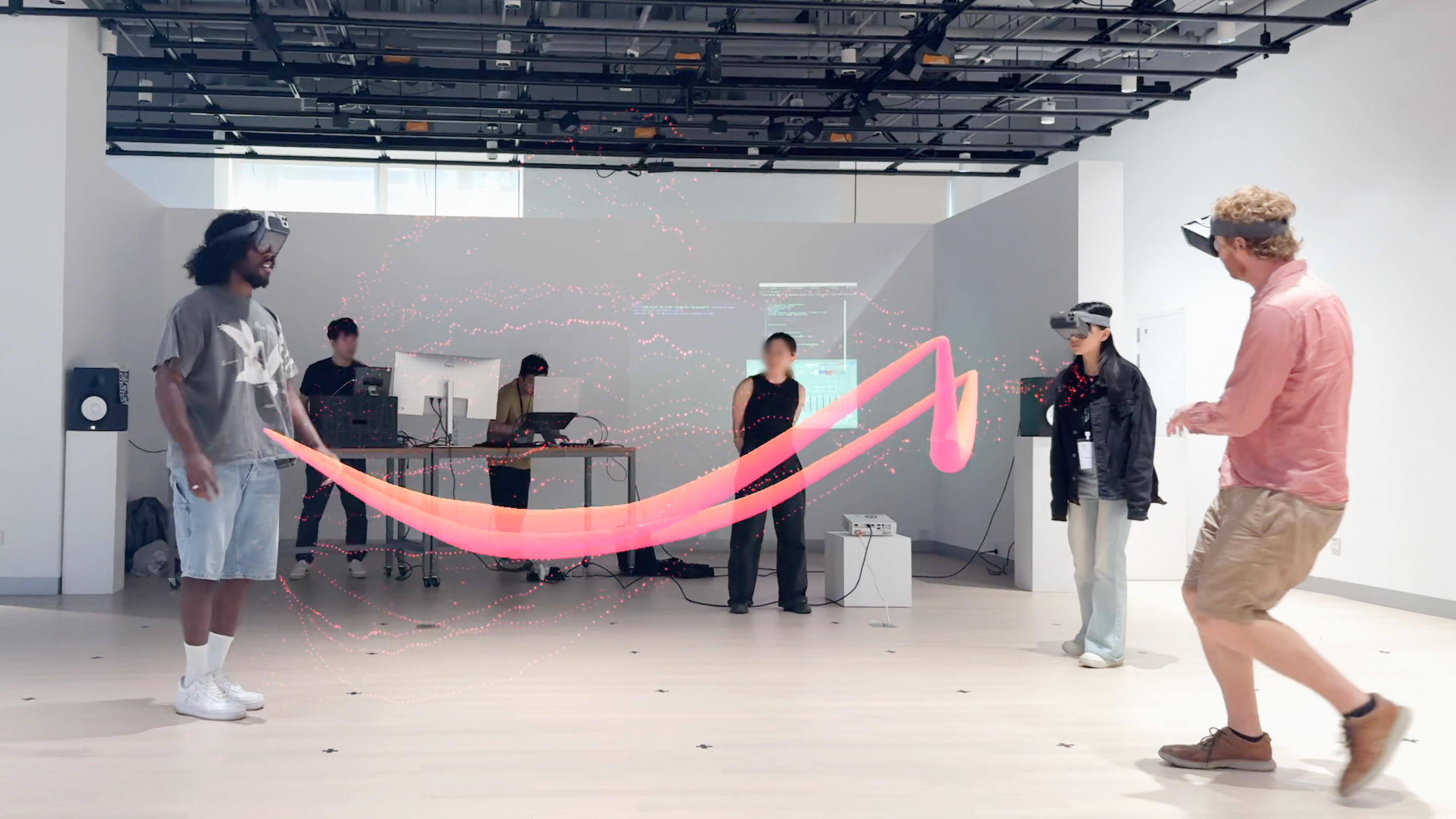}
    \caption{GravField is an experimental intercorporeal performance system for co-located mixed reality environments. Participants wear MR headsets and collaboratively improvise through body movements, altering auditory and augmented visual feedback associated with mixed reality objects (MROs). An ``Object Jockey'' (OJ) dynamically configures the ``digital physics'' of these objects on-the-fly, deciding how the MROs nudge the group's movement—much like a DJ orchestrates music tracks to motivate people to dance. Shown here is an augmented reality spectator view of an actual GravField performance: three participants swing a shared virtual ``Rope'' MRO while OJs and audience observe from the background.}
    \label{fig:teaser}
    \Description{Photograph of three participants wearing mixed reality headsets in a dimly lit performance space, captured from a spectator iPad in augmented reality. A glowing pink virtual rope visually connects the participants, curving and swaying dynamically as they swing it together. In the background, Object Jockeys seated at laptops and audience members observe the performance.}
\end{teaserfigure}

\maketitle

\section{Introduction}
How can virtual objects shape our movements without scripting our actions or overriding our agency? At the intersection of \emph{movement computing}~\cite{Biele2022Human} and \emph{extended reality} (XR), prior work often sits at two extremes. On the one hand, motion guidance systems provide prescriptive feedback, directing users toward target movements with corrective cues~\cite{Diller2024Visual,Elsayed2022Understanding,Yu2024Design}. While effective for motor learning, this approach frames unprescribed motion as error, thereby constraining exploration~\cite{Biele2022Human}. On the other hand, wearable actuation can treat the body like a puppet~\cite{Saini_Pneuma_2024,Li_Huang_Patibanda_Mueller_2023,Faltaous2024Understanding}, for example by using electrical muscle stimulation (EMS) to directly control limbs or simulate forces~\cite{Lopes2022Editorial,Patibanda2024SharedFusion,Tamaki2011PossessedHand,Lopes2017Providing,Lopes2015Impacto}. Such puppet-like control can achieve high levels of compliance and even improve performance, but it reduces agency and authorship of movement~\cite{Shahu2022Would}. Between prescriptive guidance and puppeteering lies an alternative path: \emph{nudging}, which can shape action through suggestion rather than control.

Originating in behavioral economics, a \emph{nudge} is a subtle environmental alteration that predictably influences behavior without forbidding options or significantly changing economic incentives~\cite{Thaler2008Nudge}. It functions as a strategic deployment of affordances that leverages cognitive biases non-coercively~\cite{DeRidder2023Nudgeability}. Translated to HCI as digital nudging~\cite{Weinmann2016Digital,Valta2025Digital}, this concept includes mechanisms like defaults and gentle visual cues~\cite{Caraban201923a}. In embodied settings, spatial nudges use situated cues to shape movement~\cite{Grisiute2024Spatial}. This logic extends into mixed reality (MR) \cite{Williams2025Nudging}, where virtual elements embedded in the real world can steer trajectories, for example, a virtual human figure can alter pedestrian paths through instinctive collision avoidance~\cite{Kasahara2025MR}. Although this work demonstrates that MR objects (MROs) can nudge individuals, their potential to shape collective movements is underexplored, particularly in creative contexts. In domains like interactive dance and social play, designers aim to facilitate group dynamics while preserving improvisation and creative plurality~\cite{Mueller2017Designing}. This raises two questions:




\begin{itemize}
    \item \textbf{RQ1} (Nudging Group): How can MR objects nudge intercorporeal (i.e., between-bodies) movement and collective behavior among co-located participants?
    \item \textbf{RQ2} (Live Nudging): How can non-coercive, reconfigurable MR objects (i.e., ``live nudges'') sustain exploratory participation and preserve open-endedness in group interaction?
\end{itemize}

To address these questions, we developed \textit{GravField}, a co-located MR performance system designed around live-configurable, movement-nudging virtual objects. Multiple participants, each wearing an optical see-through MR headset, share a physical space. A facilitator, or \emph{Object Jockey} (OJ), uses the system to introduce and manipulate virtual objects—such as elastic springs connecting two people, shareable ropes, or magnetic-like particle fields. The OJ fine-tunes their digital physics (e.g., elasticity, weight, force strength) in real time. These objects act as metaphorical affordances: a spring suggests tension and release, a rope invites swinging, and a magnetic field implies attraction or repulsion. Participants perceive these intangible objects through rich multisensory feedback, including visual overlays depicting object dynamics and synesthetic audio effects (e.g., a rope’s “twang” rises in pitch with tension). This design leverages embodied instincts to nudge the group toward emergent coordination patterns, while the OJ’s ability to defamiliarize the physics sustains exploration and prevents interactions from becoming routine.

We investigated GravField through a performance-led, research-through-design methodology~\cite{Benford2013Performance,Zimmerman2007Research}, conducting workshops with 25 participants. To understand the felt, bodily experience of the interactions, we captured behavioral video and conducted post-session soma-trajectory interviews~\cite{Tennent2021Articulating}. Our bricolage analysis~\cite{Rogers2015Contextualizing} of this data reveals how MR object nudges catalyze emergent group negotiation and coordination, such as the spontaneous adoption of roles and shared timings. It further shows how designed ambiguity, coupled with live reconfiguration, is key to sustaining open-ended, non-prescriptive engagement. From this, we propose \emph{MR Live Nudges} as a concept for co-located MR and describe the \emph{MR-mediated intercorporeal perception--action loop} as an analytical model of how they operate: parameterized virtual dynamics become felt cues, these cues prompt joint physical adjustments, and these adjustments feed back into how the objects are perceived and manipulated—creating a closed loop that guides collective movement without scripting it. Our work makes the following contributions to HCI~\cite{Wobbrock2016hcicontributions}:  

\begin{itemize}
\item \textbf{Artifact Contribution:} We contribute \textit{GravField}, an open-source MR performance system for live-configurable virtual objects that enables real-time nudging of group movement\footnote{The source code is available at \url{https://github.com/realitydeslab/gravfield}.}. For researchers, artists, and developers in HCI~\cite{MarquezBorbon20172015,Schacher2018What}, this system serves as a platform for studies, a tool for interactive performances, and a technical base for future systems. 

\item \textbf{Empirical Contribution:} We provide an empirical qualitative analysis of co-located MR performances, showing how virtual objects nudge collective, intercorporeal movement and nurture emergent meaning. For HCI researchers, these results explain how virtual cues shape social behavior and offer a grounded basis for future studies of technology-mediated group interaction.

\item \textbf{Theoretical Contribution:} We present the concept of \emph{MR Live Nudges}, grounded in the \emph{MR-mediated intercorporeal perception--action loop} as an analytical model of how live-configured MR objects mediate collective movement, from which we derive six actionable design strategies for non-prescriptive nudging. This work offers researchers, designers, and creative practitioners a lens for analyzing social interaction in MR and strategies for developing future systems that support group movement while preserving users' agency and expression.
\end{itemize}

\section{Background}

\subsection{Nudging, Guiding, and Human Agency}

In behavioral economics, a \emph{nudge} is an intervention where subtle changes in the ``choice architecture'' can alter people's behaviors in predictable ways without forbidding options or significantly changing economic incentives~\cite{Thaler2008Nudge}. Defaults, salience, and option ordering are canonical mechanisms that render particular actions the path of least resistance while formally preserving choice (e.g., switching organ-donation regimes from opt-in to opt-out yields large differences in consent rates~\cite{Johnson2003Defaults}). While nudges preserve \emph{autonomous choice}~\cite{DeRidder2024Simple}, users often remain unaware of the nudge, which operates on an unconscious level~\cite{Wachner2020How}.

Nudges often operate via \emph{affordances}: we can view a nudge as a strategic arrangement of affordances to influence choice~\cite{DeRidder2023Nudgeability}. \emph{Affordances}, originating from Gibson's ecological psychology~\cite{Gibson2015ecological}, describe the actionable possibilities that an environment or object naturally presents to a perceiver, relative to that perceiver's capabilities. \citet{Norman2013design} later operationalized this for design theory by making affordances perceivable through signifiers to guide intended use (e.g., door handle shaped to be grabbed signals ``pull''). Affordances themselves are neutral invitations, while a nudge adds designer intention. Nudges leverage cognitive biases in a non-coercive manner, making certain affordances stand out and turning the preferred action into the path of least resistance.

HCI researchers have adopted nudge as \textit{digital nudging}~\cite{Weinmann2016Digital, Bergram2022Digital, Valta2025Digital}, integrating nudges into user interfaces in digital systems that steer decisions while preserving voluntariness of choice~\cite{Schneider2018Digital}. \citet{Caraban201923a} systematically identified 23 distinct technology-mediated nudging mechanisms, including gentle visual cues, default settings, and framing effects.
In embodied settings, researchers refer to \emph{spatial nudges}~\cite{Grisiute2024Spatial,Duives2025Nudging}: subtle interventions that shape movement through strategically placed cues in the physical environment. For example, in transportation, the strategic placement of signs with vibrant colors on bike lanes is considered a nudge that guides people's movement choices to avoid occupying the lanes. Expanding to spatial computing settings, \emph{mixed reality nudging}~\cite{Williams2025Nudging,Kasahara2025MR} refers to interventions where virtual objects leverage embodied habits to bias movement without force. \citet{Kasahara2025MR} demonstrated this concept by positioning a transparent virtual humanoid in front of an elevator landing. This simple intervention effectively induced hesitation and detours: over half of passersby delayed or rerouted their path despite the avatar being intangible and passable.

In contrast to non-coercive nudging, a more prescriptive approach to shaping bodily movement appears in \emph{motion guidance} systems~\cite{Diller2024Visual, Elsayed2022Understanding}. These systems provide cues and corrections to direct users toward target movements~\cite{Kritopoulou2016Design}. They typically combine feedforward cues (what to do)~\cite{Muresan2023Using} with corrective feedback cues (how actual performance deviates)~\cite{Yu2024Design} to scaffold precise execution. \citet{Diller2024Visual} conducted a survey of visual cues in MR movement, including ghosted silhouettes for pose matching, joint-specific coloration for misalignment, arrow-like vector fields for direction, and rhythmic beeps for timing. This prescriptive approach can improve movement learning~\cite{TurmoVidal2023Intercorporeal} or motor skill acquisition~\cite{Sigrist2012Augmented}, yet it inherently limits exploration by discouraging any unprescribed motion as a mistake. At the other extreme, researchers have explored \emph{puppeteering} users' bodies via wearable actuation, such as using \emph{electrical muscle stimulation} (EMS) pulses~\cite{Lopes2022Editorial,Patibanda2024SharedFusion,Faltaous2024Understanding,Choudhary2025Adaptive} to actuate body parts~\cite{Tamaki2011PossessedHand}, or to simulate impact and forces~\cite{Lopes2017Providing, Lopes2015Impacto} in VR, or even affect fine motor control (e.g., finger actuation in~\cite{Tamaki2011PossessedHand}). Such direct actuation achieves compliance and even performance gains, but at a cost to user agency and authorship of movement~\cite{Shahu2022Would}.   


Across this spectrum of interventions---from subtle nudges, to explicit guidance, to direct puppeteering---researchers are fundamentally grappling with how human agency is redistributed in human--computer interaction~\cite{Chambon2014action}. \textit{The sense of agency} refers to the experience of initiating one's own voluntary actions and influencing the external world through them~\cite{Limerick2014experiencea,Cornelio2022sense}. Technology can mediate this sense: beyond just enabling individual action, interactive systems can also shape \textit{collective agency}~\cite{Bigham2014HumanComputer,Kuijer2018Coperformance,Sathya2025Cybernetic}. Agency is closely tied to user autonomy, understood as the ability to make intentional, meaningful choices within a system's constraints~\cite{Bennett2023Howa}. There are long-standing debates in HCI about how design can respect or undermine users' autonomy~\cite{Calvo2014Autonomy}---for example, discussions of persuasive technology and ``dark patterns'' often center on whether an intervention is manipulative or empowering~\cite{Alberts2024Computers,Meinhardt2025Mind}. \citet{Bennett2023Howa} reviewed HCI research on autonomy and agency, finding that conceptions of agency in our field oscillate between seeing agency as an individual's control over tools, as a property that is distributed across human--machine assemblages, and as a person's felt sense of authorship and responsibility for their actions. Following this lineage, we frame nudging not as a way to covertly override users' agency, but as a means to structure a relational field of possibilities in which participants can still improvise, refuse, and reconfigure the situation together with the system and each other.

Critics have noted that nudges can become manipulative when scaled algorithmically~\cite{Yeung2017Hypernudge} or when they exploit cognitive vulnerabilities rather than supporting informed choice~\cite{Susser2019Online}.

\subsection{Movement Computing and Soma Design} 

Movement computing (MOCO)~\cite{Schacher2018What} treats bodily motion as a primary medium for interaction and artistic expression, combining sensing~\cite{hansen2014materializing}, analysis, and movement theory across domains such as dance~\cite{Zhou2021Dance}, games~\cite{Mueller2014Movementbased}, sports~\cite{Mueller2011Designinga}, and rehabilitation. In HCI, movement-based design~\cite{Loke2013Moving,VanRheden2024Why,wildeMoveDesignDesign2011,Hummels2007Move} explores how interactive systems can foreground expressive, qualitative movement rather than only efficient task execution. Examples include exertion games~\cite{Mueller2016Exertion} and bodily play~\cite{Mueller2018Experiencinga}, which treat physical effort as material for social bonding and affective experience~\cite{Isbister2015Guidelinesb}. Movement computing also includes rich work on movement—sound interaction~\cite{Schacher2016Moving,Nabi2024Embodied,Gulino2024Sounding,Reed2024Sonic,Potapov2025Movement}, where movement becomes an instrument for real-time sonic expression~\cite{Alaoui2015Interactive}. For example, the CO/DA live-coding environment lets a performer write code that maps dancers' movements to sound in real time, treating choreographic improvisation and live coding as a joint practice~\cite{Francoise2022CO}.

MOCO is grounded in the ``somatic turn''~\cite{Loke2018Somatic} in HCI, which treats the body as subject rather than object of design. Somaesthetic interaction design~\cite{Hook2017SomaBased} draws on~\citet{Shusterman1999Somaesthetics}'s somaesthetics to emphasize first-person felt experience, cultivating bodily awareness and aesthetic appreciation rather than merely optimizing efficiency or accuracy~\cite{Hook2016Somaesthetic}. Soma design~\cite{MartinezAvila2020Soma, Hook2026Change,Hook2018Designing, Hook2020Soma} operationalizes this stance through sustained bodily practices, sensitizing exercises, and rich material exploration, encouraging designers to iteratively feel, reflect, and redesign~\cite{Hook2021Unpacking}. Within this lineage, \emph{movement qualities} such as heaviness, fluidity, or tension become central constructs; interactive systems are designed to sense and reflect these qualities back to users rather than only measuring kinematic correctness~\cite{Alaoui2012Movement, Mentis2013Seeing, Alaoui2015Interactive, FdiliAlaoui2017Seeing, TurmoVidal2024Body, TurmoVidal2024BodySensations}.

Methodologically, soma design has introduced tools for tracing the temporal unfolding of embodied experience. Soma trajectories~\cite{Tennent2021Articulating} document changes in engagement, breakdown, and re-attunement over time, providing an analytic lens on how participants move through phases of exploration, learning, and sense-making~\cite{Benford2025Tangles}. Soma-based ideation methods have designers use their own bodies to explore scenarios and constraints. Bodystorming~\cite{MarquezSegura2016Bodystorming,schleicher2010bodystorming} involves full-body enactment of scenarios to reveal situated constraints and opportunities. Embodied sketching~\cite{MarquezSegura2016Embodied} similarly relies on improvised movement, sometimes supported by minimal props or projections, to prototype interaction dynamics before committing to technical implementation. In our work, we draw on movement computing and soma design as both conceptual and methodological foundations: GravField was developed through soma design and movement-based design processes—bodystorming and embodied sketching—and later analyzed using soma trajectories, aiming to understand not only what movements occurred but how they felt and evolved over time.

\subsection{Improvisational Dance and Intercorporeal Design}

Improvisational dance—the art of spontaneous creation that unfolds moment-to-moment without a scripted outcome—offers a rich model for open-ended, relational movement. Practitioners continuously sense, adapt, and create anew, where practitioners exercise agency through openness and spontaneity, allowing improvisation to produce a stream of novel, unrepeatable moments highly valued in artistic creativity~\cite{Ravn2020Investigating}. Contact Improvisation provides a paradigmatic example: dancers maintain points of contact, share weight, and continuously negotiate balance to generate trajectories without fixed choreography~\cite{Paxton1975Contact,Pallant2006Contact}. \citet{Kimmel2018Sources} show how such practices enable open-ended embodied creativity to emerge through minimal scores and responsive partnering rather than top-down instruction. Phenomenological and enactive accounts describe intercorporeality as the way bodies co-constitute a shared world—each action simultaneously serving as stimulus and response for others through kinesthetic empathy, mutual incorporation, and participatory sense-making~\cite{Fuchs2009Enactive, Fuchs2016Intercorporeality, Tanaka2015Intercorporeality,DeJaegher2007Participatory,daSilva2024How,Gordon2025Interpersonal,Shen2025Haptic}.

HCI has begun to work explicitly with such improvisational and intercorporeal practices. Soma-oriented projects like Drone Chi~\cite{LaDelfa2020Drone} and Dancing with Drones~\cite{Dong2024Dances} explore how human--drone couplings can be tuned so that the drone becomes an expressive partner rather than a tool. Intercorporeal biofeedback systems reflect bodily signals across people—for example, sharing breathing or heart rate—to scaffold joint attention and coordinated movement learning~\cite{TurmoVidal2023Intercorporeal,Patibanda2024SharedFusion}. Co-creative dance systems such as Cyber Subin~\cite{Pataranutaporn2024HumanAI} and LuminAI~\cite{Trajkova2024Exploring} treat AI agents as improvisational partners, using mirroring, turn-taking and riffing to lead participants beyond habitual patterns while maintaining co-authorship. These works show how technology can mediate intercorporeal relations without collapsing them into either pure control or pure autonomy.

The notion of \emph{intercorporeal design} proposed by~\citet{Stepanova2024Intercorporealb} explicitly argues for dissolving self--other dualisms, treating interaction as emerging from continuous bodily coupling rather than discrete individuals. They call for design methods that attend to mutual vulnerability, shared rhythms, and shifting agency in more-than-human assemblages. Building on these perspectives, we treat GravField as an intercorporeal design for improvisational dance: mixed reality objects—metaphors such as springs, ropes, and fields—serve as shared mediators linking multiple bodies and shaping intercorporeal movements.

\subsection{Live Coding, Algoraves, and NIME}

Rave culture and DJ-mediated dance events provide a complementary lineage for thinking about technologically orchestrated collective movement. Raves have been analyzed as techno-cultural and spiritual practices where sound, light and crowd dynamics co-produce altered states and communal affect~\cite{StJohn2004Rave}. Within this setting, the DJ acts as a live mediator who sequences tracks, manipulates tempo and texture, and reads the crowd, effectively ``programming'' waves of intensity that shape how people move together~\cite{Gates2006DJs}. However, as the title of \textit{``How Do You Know He's Not Playing Pac-Man While He's Supposed to Be DJing?''}~\cite{Montano2010How} suggests, if a DJ delegates agency to the autonomous system, the audience may not be able to tell. 

Algoraves~\cite{collins2014algorave} extend this logic by making algorithms themselves the performance medium. In musical live coding, performers write and modify code in real time to generate sound, often projecting the code so that the audience can witness the process~\cite{Collins2003Live}. \citet{Blackwell2022Livea} describe the live coding community’s emphasis on improvisation, transparency, and open-ended configuration, framing code as a malleable instrument that can be reshaped mid-performance. \citet{Xambo2024Human} discussed human--machine agencies in live coding for music performance. This ethos inspires our Object Jockey, who live-configures mappings between mixed reality objects, digital physics, and sound rather than playing fixed tracks.

In parallel, the NIME (New Interfaces for Musical Expression) community~\cite{MarquezBorbon20172015} contributes a rich lineage of bodily and movement-based instruments that resonates with our focus on collective movement and programmable media. A series of projects explicitly treats the performer’s ``\textit{body as the instrument}'': gestural interfaces that foreground full-body performance~\cite{Mainsbridge2014Body}, installations where a blindfolded performer’s body is ``played'' by visitors while motion sensors drive sound~\cite{Bomba2019Somacoustics}, and soma-informed instruments designed around balance, tension, and first-person felt experience~\cite{MartinezAvila2020Soma}. Together, these contributions frame digital instruments as deeply embodied and often distributed across the performer’s entire moving body. At the same time, NIME offers clear precedents for multi-user and multi-body instruments in which several performers jointly play a single system~\cite{Aylward2006Sensemble,Feng2024Codesigning,Abraham2024Choreographing}. 
\citet{Jorda2005MultiUser}’s analysis of multi-user instruments distinguishes different models of shared control and illustrates how instruments such as reacTable are designed to foster interdependent, collaborative performance,  while~\citet{Rotondo2012ManyPerson} introduce \emph{``many-person instruments''} that require multiple participants to act together for the instrument to function as intended. Ensemble Feedback Instruments distribute interconnected audio processes across multiple performers’ stations so that the ensemble effectively becomes one coupled feedback system~\cite{Rosli2015Ensemble}, \citet{Erdem2019Vrengt} proposes a shared body--machine instrument for music--dance performance that links dancer and musician through EMG, breathing, and sound, and the notion of Digital Dance and Music Instruments argues for instruments that are intrinsically defined by the coupling of moving bodies and sound production~\cite{Tragtenberg2019Concept}. This ensemble-focused perspective is further supported by infrastructures such as the Stanford Laptop Orchestra~\cite{wang2009stanford} and community reflections on NIME, which highlight ensemble configurations and instrument ecologies as key design concerns~\cite{MarquezBorbon20172015}.  Within this lineage, GravField can be read as a \emph{``live-coded intercorporeal instrument''}: an Object Jockey configures MR objects and their physics as an instrument that is literally played through multiple bodies moving in a shared space.

\subsection{Co-located Mixed Reality and Materiality}

Mixed reality head-mounted displays overlay digital content onto the physical environment while preserving visibility of bodies, eye contact, and gesture, enabling situated social interaction alongside virtual elements~\cite{Speicher2019What}. Early multi-user AR systems already emphasized this “see-through co-presence,” where multiple people viewed spatially aligned graphics from their own viewpoints and coordinated via ordinary social cues~\cite{BillinghurstCollaborative}. More recently, work on programmable reality has highlighted how virtual properties such as gravity, friction, or field strength can be dynamically reconfigured to subtly reshape sensorimotor contingencies and everyday behavior, treating MR environments as configurable fields of forces rather than fixed overlays~\cite{Suzuki2025Programmable}.

Within this landscape, collaborative MR describes systems that allow multiple users to share and act upon the same virtual content, either remotely or in face-to-face co-located settings~\cite{Billinghurst2002Collaborativea,Schafer2023Survey}. Co-located MR~\cite{McGill2020,Bhattacharyya2019Brick,Sereno2020Collaborative,Dagan2022Projecta,Lin2024Cell,Miedema2019Superhuman,Grandi2019Characterizinga,Zhou2019Astaire,Radu2021Survey,Thomas2012survey,He2019Exploring,DSouza2018Augmentingc,Reig2023Supporting, Hu2023MOFA} adds shared spatial alignment so that all participants perceive consistent placement and motion of virtual objects, enabling joint attention and coordinated action around anchors in the physical space. This requires robust spatial anchor sharing and drift compensation across heterogeneous devices; for example, InstantCopresence~\cite{hu2023InstantCopresence} demonstrates lightweight methodologies for synchronizing coordinate frames between handheld and headworn AR to support spontaneous co-located multiplayer play, showing how co-located MR experiences can involve both HMD and non-HMD participants while mediating intercorporeal awareness and collective sense-making. 

MR is also an inherently embodied medium~\cite{McVeighSchultz2024Designing}: head and hand tracking, and spatial audio invite people to move, reach, and orient their bodies relative to virtual entities, while co-present others simultaneously negotiate proxemics, facing, and turn-taking. Co-located MR interaction has been used to reorganize social spacing and shared attention through force-like cues and spatialized feedback, illustrating how virtual entities can act as attractors or resistive fields for group movement~\cite{Kim2022Weight}. These affordances make MR suitable for movement-centered and intercorporeal experiences, but they also foreground questions of agency and control~\cite{Suzuki2025Programmable}: who configures the shared environment, and how is this influence perceived and negotiated by participants?  

Despite these possibilities, the materiality of MR objects remains a challenge. Virtual objects are intangible; users cannot directly feel their weight, texture, or resistance. One approach introduces dedicated haptic hardware to convey forces and contact~\cite{Lopes2017Providing}, but such solutions can be costly or impractical in public, large-scale, or performance settings. As a result, designers often turn to synesthetic and multimodal strategies that use visual exaggeration and sound as primary channels for conveying force and material properties. Research on synesthetic design~\cite{Haverkamp2013Synesthetic,DesnoyersStewart2025Synedelica} shows how tightly coupled audiovisual mappings can evoke impressions of weight, tension, impact, and texture without physical haptics~\cite{Merter2017Synesthetic,Sigrist2012Augmented}. HCI and VR work further demonstrates that carefully coordinated movement, sound, and imagery can induce audio--haptic and visuo--haptic illusions of stiffness, heaviness, and surface qualities~\cite{Kim2022Weight,Outram2016Extranormal}. In this sense, many MR experiences rely on “as-if haptics” where sound and visuals simulate the presence of forces, allowing users to treat virtual springs, fields, and impacts as materially meaningful despite the absence of direct physical contact.

\section{Creating GravField}

\subsection{Design Rationale and Goals}
GravField emerged through a research-through-design~\cite{Zimmerman2007Research} trajectory in which system building, embodied studio rehearsals, and inquiry were tightly intertwined. Rather than specifying a fixed design upfront, we treated movement itself as the primary design material---iteratively prototyping, improvising with our own bodies~\cite{MarquezSegura2016Bodystorming,schleicher2010bodystorming,MarquezSegura2016Embodied}, and reflecting on how different configurations of MR object feedback reshaped the way people moved together.

GravField is inspired by two movement practices. In Contact Improvisation (CI)~\cite{Paxton1975Contact, Pallant2006Contact}, partners continuously negotiate touch, weight, and shared momentum, organizing collective movement without predetermined choreography. William Forsythe's ``\emph{choreographic objects}''~\cite{forsythe2011choreographic} showed that everyday spatial constructs can reshape how bodies relate and move together (see Appendix~\ref{sec:ci_inspiration} for the full design narrative). GravField translates both insights into mixed reality through \textit{live-configurable} Mixed Reality Objects (MROs) whose ``digital physics'' (e.g., mass, tension, polarity) can be tuned during performance to gently reshape participants' perception--action loops. From an iterative process, we converged on three design goals:

\begin{itemize}
  \item \textit{Facilitating Emergent Group Movement.} Support open-ended, non-goal-oriented collective exploration in which relational dynamics emerge from bodily negotiation rather than instruction or task completion.
  \item \textit{Enabling Live Nudging.} Empower a human facilitator---the Object Jockey (OJ)---to responsively reconfigure object affordances in real time, gently steering the trajectory and energy of collective movement without scripting outcomes.
  \item \textit{Perceptible and Responsive ``Digital Physics.''} Create tight perceptual coupling between participant action and audiovisual feedback so that object properties (e.g., tension, weight, attraction) can be \textit{felt} through synesthetic perception, while retaining enough ambiguity for participants to discover unexpected behaviors.
\end{itemize}

\subsection{Research team and roles}
Our interdisciplinary team combined MR design, HCI, dance practice, and live performance:
\begin{itemize}
  \item \textbf{R1:} first author (MR researcher and MR designer).
  \item \textbf{R2:} MR designer and HCI researcher.
  \item \textbf{D1:} dancer with long-term Contact Improvisation experience.
  \item \textbf{D2:} new media artist with dance and performance practices.
  \item \textbf{OJ1:} sound artist and live coder (focus on orchestration and tuning “how it feels” through sound and parameter changes).
  \item \textbf{OJ2:} new media artist and MR engineer (focus on MR implementation, OSC routing, and sound mapping into Ableton Live).
\end{itemize}
We met through a sequence of informal studio rehearsals in a university design-studio stage, each lasting 4--5 hours. These rehearsals functioned as internal trials for embodied sketching, rapid prototyping, and reflection, allowing us to iterate before deploying the system in formal sessions.

\subsection{Four-stage iterative development}
\label{sec:four_stage_body}
We developed GravField through four iterative stages: (1)~single-player, sound-only embodied sketching to establish synesthetic action--sound loops; (2)~multi-player, sound-only sessions that revealed intercorporeal signals as the most meaningful design material; (3)~adding MR visuals, where physical-object metaphors act as productive constraints on movement; and (4)~adding live OJ configuration, which transformed GravField from a fixed artifact into a performable system. Through this process we converged on three MROs---Rope, Spring, and Magnetic Field---and an OSC-based orchestration architecture. Appendix~\ref{sec:design_iterations} details each stage's goals, prototypes, and key reflections.

\subsection{System Architecture and Implementation}
\label{sec:implementation}

\begin{figure*}[ht]
  \centering
  \includegraphics[width=\linewidth]{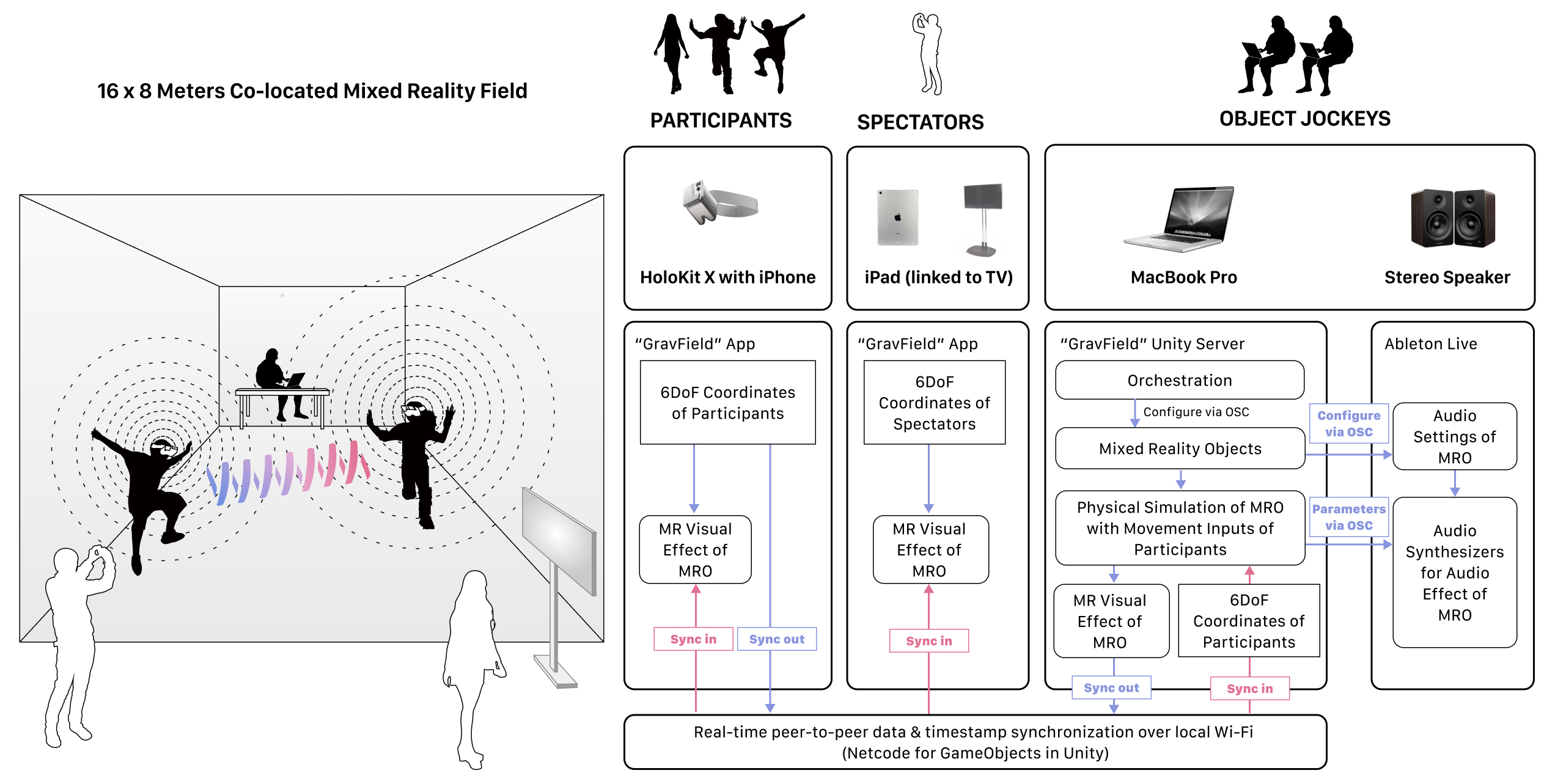}
  \caption{System architecture of GravField. HoloKit X headsets stream six-degrees-of-freedom pose data to a Unity server, which computes Mixed Reality Object physics, renders visual effects back to headsets, and sends Open Sound Control messages to Ableton Live for real-time audio synthesis. A spectator iPad and TV display the mixed reality overlay for audience viewing.}
  \label{fig:system}
  \Description{Block diagram with five labeled component boxes arranged left to right. On the far left, multiple HoloKit X headset icons with iPhones represent participants' devices. A rightward arrow labeled six-degrees-of-freedom pose data connects them to a central MacBook Pro box labeled GravField Unity Server, which computes mixed reality object physics. A leftward arrow labeled mixed reality visual effects returns rendered graphics to the headsets. From the server, a rightward arrow labeled Open Sound Control messages leads to an Ableton Live box on the same laptop. From Ableton Live, a rightward arrow labeled synthesized audio connects to a stereo speakers icon. A separate downward arrow from the server labeled mixed reality overlay stream connects to a spectator iPad and a television at the bottom, used for audience viewing. Each arrow is color-coded: blue for visual data, orange for audio data, and yellow for pose tracking.}
\end{figure*}

The iterative process described above converged on the following architecture (Figure~\ref{fig:system}). GravField is a co-located MR system supporting up to four participants and one or two OJs. It consists of: (1) participant MR headsets running the GravField client app, (2) a shared-coordinate co-location layer, (3) a Live Orchestration system (Unity) operated by the OJ, (4) an OSC messaging layer that links movement signals, physics parameters, and audio control, and (5) a spectator view (tablet device) for external observation.

\paragraph{Participants and MR device (6DoF head pose)}
Each participant wears a HoloKit X~\cite{Hu2024HoloKit} headset, an affordable open-source MR device that uses an iPhone for rendering and on-device tracking, making multi-player deployment practical. The system uses the participant’s 6DoF head pose (position + orientation) as the primary embodied input. We intentionally designed around head pose because it is robust, available on commodity mobile MR stacks, and preserves uninstrumented full-body movement (participants can still move naturally without holding controllers).

\paragraph{Spatially synchronized co-located MR}
To ensure all participants share a consistent spatial frame of reference, we synchronized the shared coordinate space using \emph{InstantCopresence}~\cite{hu2023InstantCopresence} technology. After initial calibration via external QR code on the ground, the system streams real-time data—including participant positions, object positions, and physics properties—between the Live Orchestration system (typically the OJ's laptop serving as the server) and participants' headsets through a low-latency local Wi-Fi network. This bidirectional synchronization is managed through Netcode for GameObjects\footnote{\url{https://github.com/Unity-Technologies/com.unity.netcode.gameobjects}}.

\begin{figure*}[ht]
  \centering
  \includegraphics[width=1\linewidth]{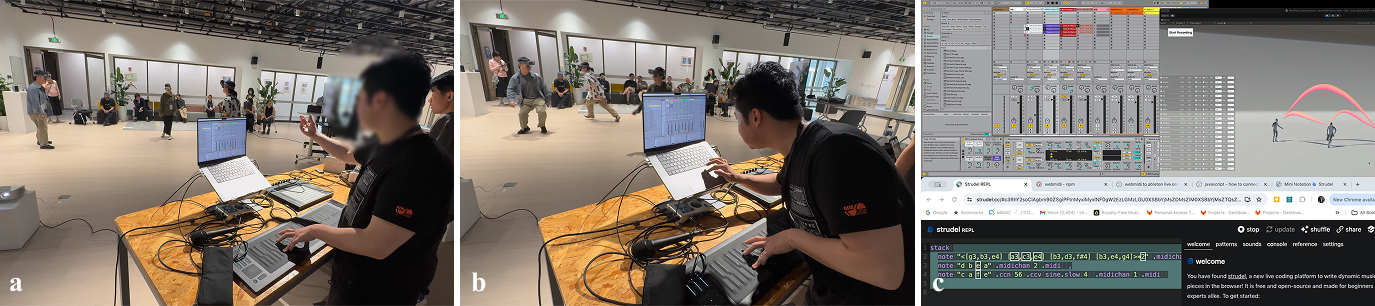}
  \includegraphics[width=1\linewidth]{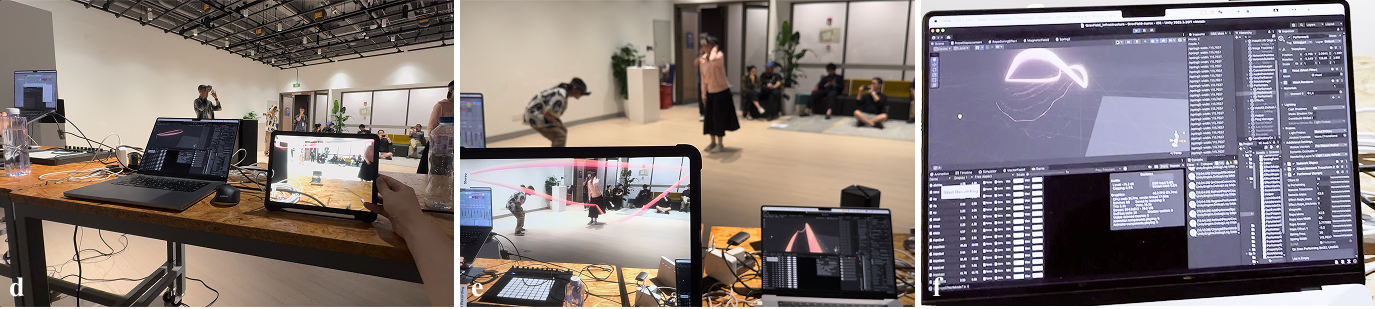}

  \caption{OJs configure mixed reality objects during the GravField workshop. (a) OJ1 communicates with participants on stage; (b) OJ1 live-configures nudging parameters on a laptop; (c) OJ1's Unity interface showing the real-time AR scene view, a live-coding panel for binding physical properties, and Ableton Live for sound synthesis; (d) OJ2 monitors the performance alongside OJ1; (e) a spectator iPad displays the live-streamed AR view of participants and MRO visuals; (f) close-up of OJ2's Unity interface with the live-streamed MR overlay.}
  \label{fig:oj1}
  \Description{Grid of six photographs showing Object Jockeys at work during the GravField workshop, arranged in two rows of three. Top row: (a) Object Jockey 1 standing beside the stage and speaking with headset-wearing participants; (b) Object Jockey 1 seated at a laptop adjusting MRO parameters via a control interface; (c) close-up of the laptop screen showing the Unity editor with a live-coding panel, a 3D scene view of the virtual rope, and Ableton Live audio tracks. Bottom row: (d) both Object Jockeys seated side by side at their respective screens; (e) an iPad propped on the desk displaying the spectator augmented reality view; (f) close-up of Object Jockey 2's Unity interface showing the live-streamed virtual rope overlaid on a camera feed of the physical space.}
\end{figure*}

\paragraph{Live Orchestration system and audio pipeline}
The Live Orchestration system is a Unity application running on the OJ's laptop that serves as the central hub of GravField (see Figure~\ref{fig:oj1}). It receives participants' head poses in the shared coordinate frame, runs the physics simulation for each active MRO (computing movement-derived variables such as distances, velocities, and accelerations), broadcasts updated object states to all participant devices for consistent rendering, and exposes live-tunable parameters (e.g., mass, elasticity, field strength) to the OJ via OSC. The OJ can spawn and manipulate virtual objects, attach them to participants, and adjust parameters in real time, thereby altering the kinesthetic sensations participants experience.

\begin{figure*}[htbp]
  \centering
  \includegraphics[width=0.85\linewidth]{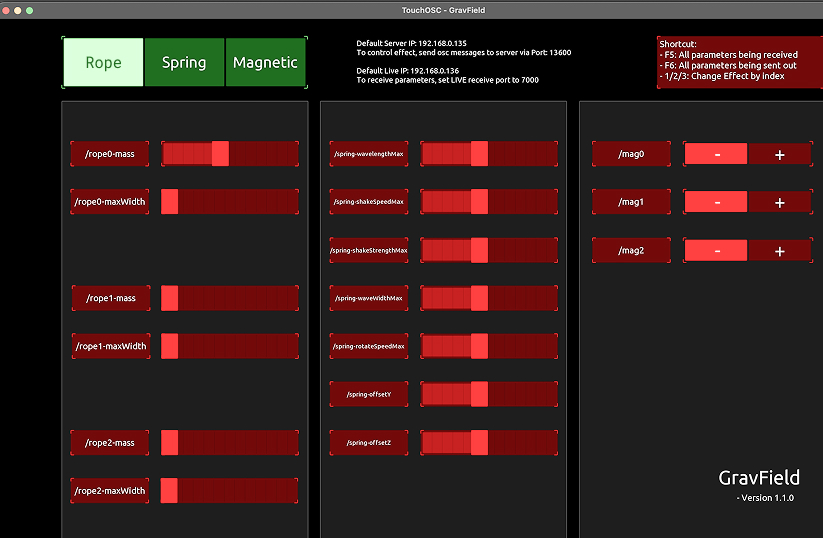}
  \caption{OJ1's TouchOSC control panel for real-time configuration of ``digital physics.'' Three mode buttons (top) switch between the Rope, Spring, and Magnetic MROs, each exposing mode-specific parameter sliders. The OJ adjusts these on-the-fly during a performance to nudge collective movement.}
  \label{fig:oj3}
  \Description{Screenshot of the Object Jockey's TouchOSC control interface displayed on a tablet. Three green buttons at the top select between Rope, Spring, and Magnetic modes. Mode-specific sliders and controls appear below: Rope mode shows mass and maximum width sliders; Spring mode shows wavelength, shake speed, wave width, and rotation controls; Magnetic mode shows plus and minus buttons for three independent field strengths. A server address and port number are shown at the bottom.}
\end{figure*}

We implement the OJ control surface using TouchOSC (see Figure~\ref{fig:oj3}). OJ1 typically manages this panel to adjust the MRO ``physics'' during the session (e.g., making a rope heavier, a spring stiffer, or a magnetic field stronger). These controls can also include higher-level actions such as spawning/activating an MRO preset or changing how participants are connected.

Movement-derived variables generated by participants' movements---such as rope velocity, spring length, and pairwise distances---are streamed as OSC (Open Sound Control)\footnote{\url{https://opensoundcontrol.stanford.edu/}} messages into Ableton Live\footnote{\url{https://www.ableton.com/en/}}. The OJ maps these data streams to sound modulators (e.g., MIDI pitch, filter cutoffs, effects), so that sound acts as a perceptual ``carrier'' of digital physics: the same variable (e.g., tension) can be simultaneously rendered visually (thicker line, brighter particles) and sonified (pitch rise, filter opening), enabling participants to perceive dynamics without haptic devices. The resulting audio plays through the MR headsets and high-amplification stereo speakers.

\paragraph{Spectator system}
Spectators can observe the MR scene via a spectator device (e.g., iPad) that is also aligned to the shared coordinate system. This feed can be displayed on external screens (TV/projection) as a fixed camera view, and handheld devices can be used to explore different viewing angles.

\begin{figure*}[htbp]
  \centering
  \includegraphics[width=0.85\linewidth]{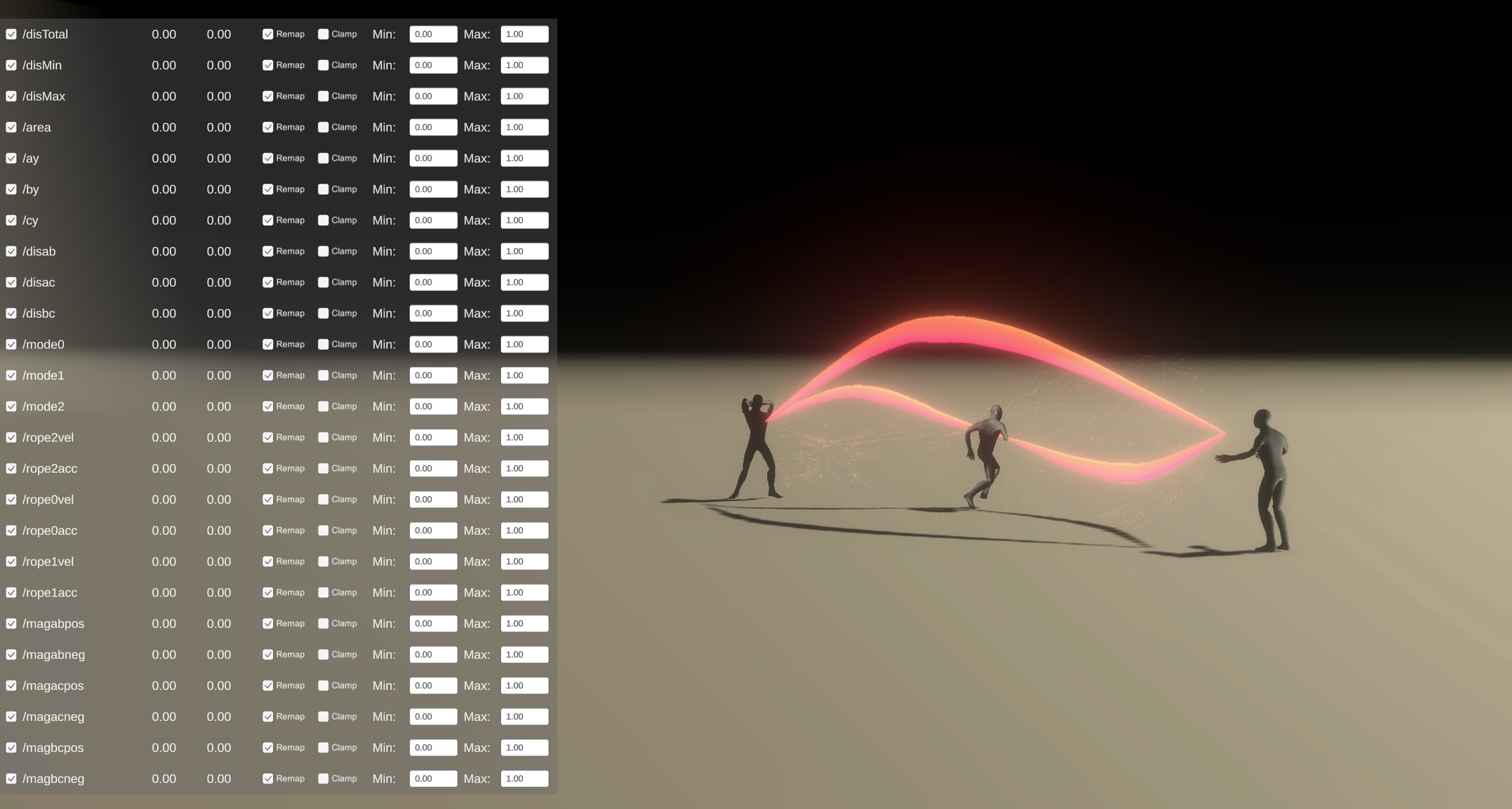}
  \caption{OJ2's Unity control interface for monitoring participants' real-time state. Left panel: incoming OSC parameters including interpersonal distances, heights, rope velocity and acceleration, and magnetic field values. Background: the simulated scene rendering three participants connected by Rope MROs with real-time data streaming.}
  \label{fig:oj2_control}
  \Description{Screenshot of OJ2's Unity control interface. The left panel displays a list of incoming OSC parameter channels with real-time values, including disTotal, disMin, disMax, area, individual heights, pairwise distances, mode selectors, rope velocity and acceleration for each participant pair, and magnetic field positive and negative values. Each parameter shows current and remapped values with min and max range settings. The background shows a simulated 3D scene rendering three participants as dark silhouettes connected by glowing orange Rope MROs in a warmly lit virtual environment, with real-time data streaming from participant headsets.}
\end{figure*}

\paragraph{OJ workflow during sessions}
In our live sessions, two OJs collaborated throughout. OJ1 (live-coding musician) typically led the interaction, while OJ2 (MR engineer) monitored and supported. They operated three complementary interfaces: (1)~a Unity scene displaying live participant data and active MROs (primarily OJ2, see Figure~\ref{fig:oj2_control}); (2)~a TouchOSC control surface with sliders, toggles, and buttons for switching MROs and adjusting their properties (both OJs); and (3)~an Ableton Live audio-remix interface receiving all movement-derived variables via OSC (primarily OJ1). During a session, the OJs observe participants' movements and energy through the spectator view, shift between MROs, adjust physical parameters, and remap movement-derived variables to sonic parameters. Through these real-time adjustments, the OJs nudge collective movement not through verbal instruction but by subtly reconfiguring the field of forces participants experience.

\subsection{Mixed Reality Objects}

\begin{figure*}[ht]
  \centering
  \includegraphics[width=1\linewidth]{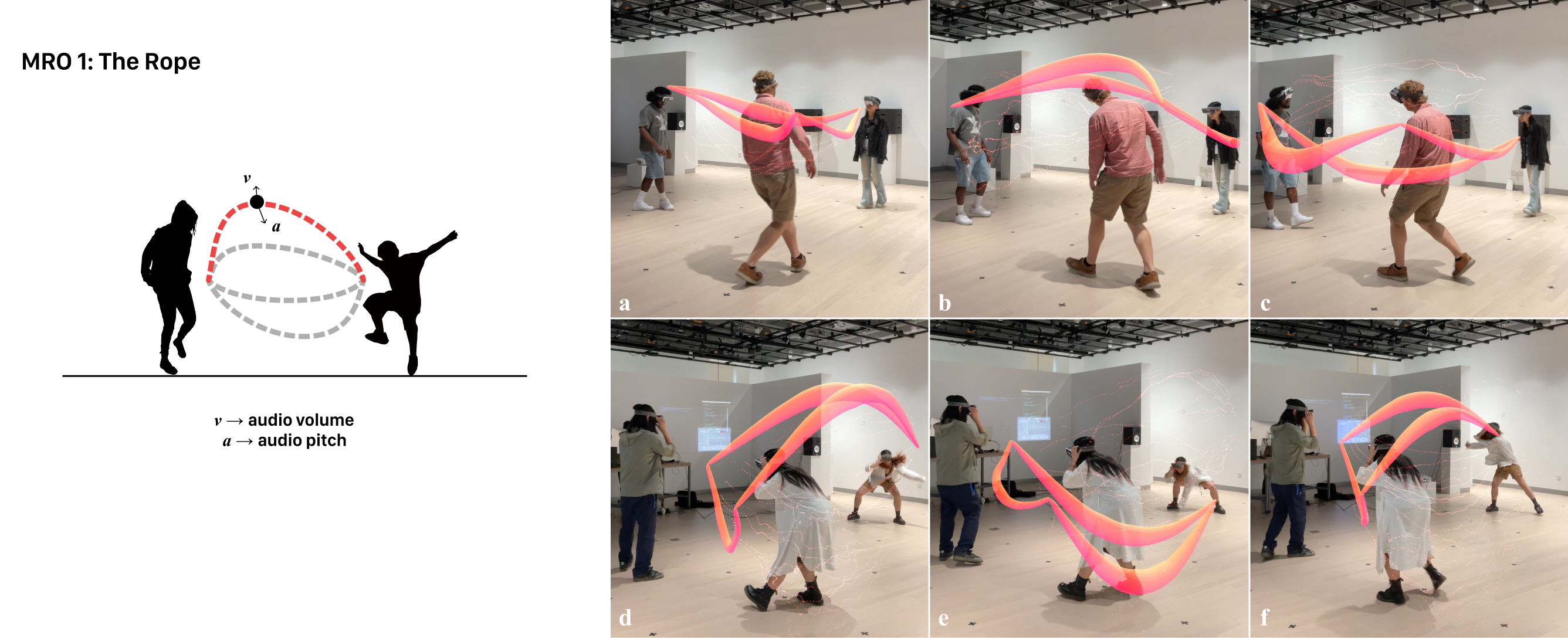}
  \caption{The ``Rope'' Mixed Reality Object. Left: rope velocity ($v$) maps to audio volume and centripetal acceleration ($a$) to pitch, rewarding faster continuous motion. Right: participants swinging the virtual rope during a performance.}
  \label{fig:rope}
  \Description{Two-panel figure of the Rope mixed reality object. Left panel: a schematic of two silhouetted figures connected by a curved virtual rope, annotated to show that rope velocity (v) controls audio volume and centripetal acceleration (a) controls pitch. Right panel: six photographs of participants wearing head-mounted displays swinging a glowing pink virtual rope through arcs and circular motions, each photo capturing a distinct movement phase.}
\end{figure*}

\begin{figure*}[ht]
  \centering
  \includegraphics[width=1\linewidth]{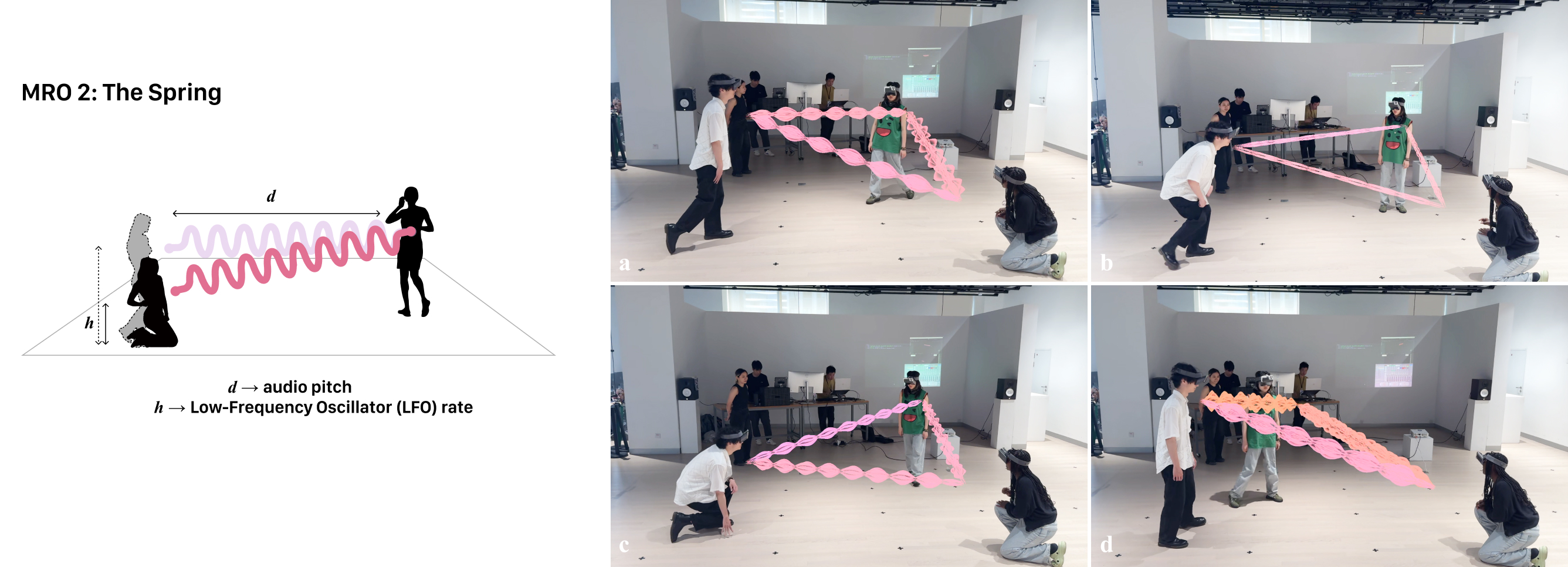}
  \caption{The ``Spring'' Mixed Reality Object. Left: horizontal distance ($d$) controls pitch and relative height ($h$) controls a low-pass filter, encouraging push--pull and see-saw dynamics. Right: participants exploring vertical and lateral movements during a performance.}
  \label{fig:spring}
  \Description{Two-panel figure of the Spring mixed reality object. Left panel: a schematic of two silhouetted figures connected by a wavy spring-like line, annotated to show that horizontal distance (d) controls audio pitch and relative height (h) controls a low-pass filter cutoff. Right panel: four photographs of participants connected by a taut, glowing pink ribbon with a sine-wave texture, showing lateral distance changes and synchronized squatting motions; the ribbon's wave amplitude increases visibly when participants vocalize or clap.}
\end{figure*}

\begin{figure*}[ht]
  \centering
  \includegraphics[width=1\linewidth]{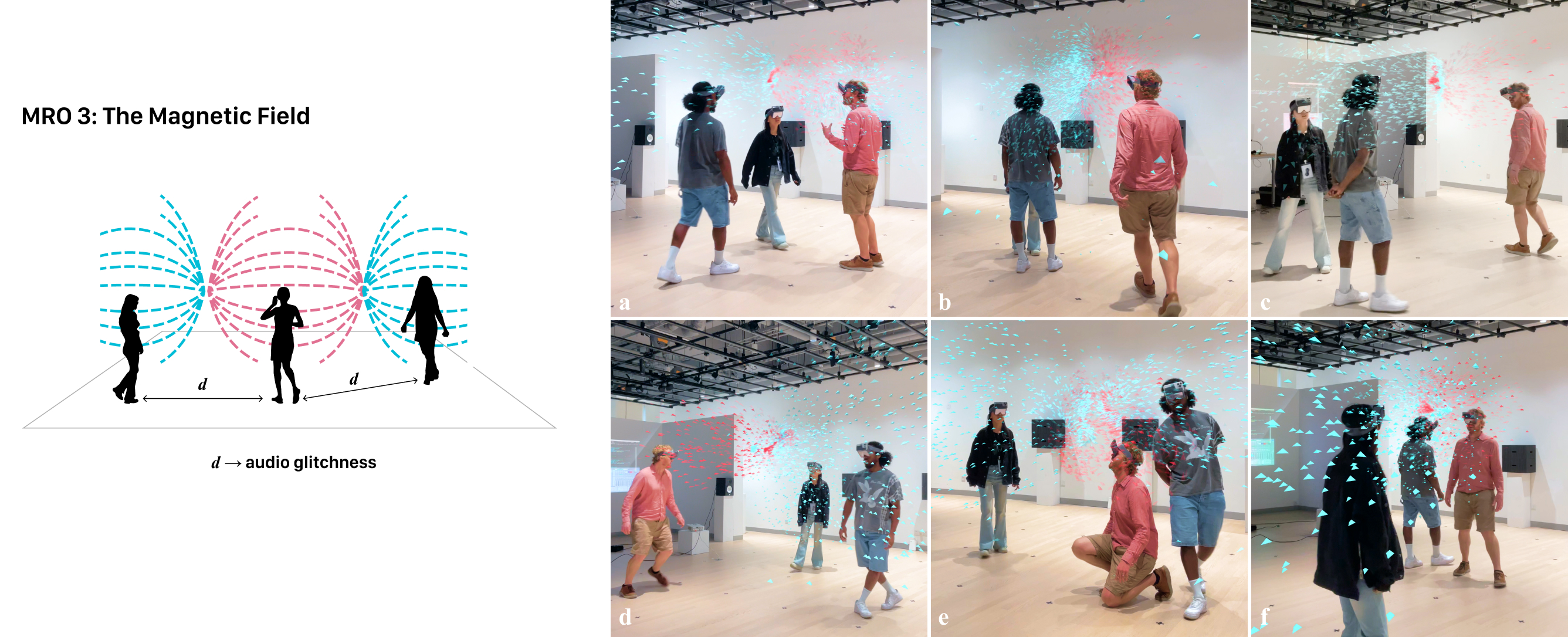}
  \caption{The ``Magnetic Field'' Mixed Reality Object. Left: inter-participant distance ($d$) drives audio glitch intensity, with polarity assignments governing attraction and repulsion. Right: participants navigating drifting particles during a performance.}
  \label{fig:magnetic}
  \Description{Two-panel figure of the Magnetic Field mixed reality object. Left panel: a schematic of three silhouetted figures with curved magnetic field lines; the outer two figures show blue fields, the middle figure shows a red field, and opposite-color fields curve toward each other to indicate attraction, with an annotation noting that inter-participant distance (d) controls audio glitch intensity. Right panel: six photographs of participants moving through scattered blue and red virtual particles radiating from each person, with particle density and audio distortion varying based on proximity and polarity assignment.}
\end{figure*}

\begin{table*}[htbp]
\centering
\begin{tabularx}{\linewidth}{lXXX}
\toprule
\textbf{MRO} & \textbf{Configurable parameters} & \textbf{Movement-derived variables} & \textbf{Sound mappings*} \\
\midrule
Rope & \begin{tabular}[t]{@{}l@{}}Mass;\\ Width\end{tabular} &
\begin{tabular}[t]{@{}l@{}}Midpoint velocity ($v$); \\ Midpoint acceleration ($a$)\end{tabular} &
\begin{tabular}[t]{@{}l@{}}$v$ $\rightarrow$ Volume (faster $\rightarrow$ louder);\\ $a$ $\rightarrow$ Pitch (sharper turns $\rightarrow$ higher)\end{tabular} \\
\midrule
Spring & \begin{tabular}[t]{@{}l@{}}Wave width;\\ Wave offset;\\ Max.\ wave length\end{tabular} &
\begin{tabular}[t]{@{}l@{}}Interpersonal distance ($d$);\\ Heights ($h$);\\ Interpersonal height difference ($\delta h$);\\ Microphone volume\end{tabular} &
\begin{tabular}[t]{@{}l@{}}$d$ $\rightarrow$ Pitch;\\ $h$ $\rightarrow$ LFO rate;\\ Mic volume $\rightarrow$ Spring width\end{tabular} \\
\midrule
Magnetic Field & \begin{tabular}[t]{@{}l@{}}Polarity (+/\textminus);\\ Field strength\end{tabular} &
\begin{tabular}[t]{@{}l@{}}Interpersonal distance ($d$);\\ Field turbulence\end{tabular} &
\begin{tabular}[t]{@{}l@{}}$d$ + turbulence $\rightarrow$ Glitchness\end{tabular} \\
\bottomrule
\end{tabularx}
\vspace{8pt}
\centering\small *Mapping relations and mapping parameters can be changed by OJs during performance.
\caption{Overview of the mixed reality objects (MROs), their configurable parameters, movement-derived variables, and sound mappings.}
\label{tab:mro}
\end{table*}

Because virtual objects cannot exert physical force or haptic sensation on a body the way physical props do, we designed each MRO as a tightly coupled audiovisual entity: its visual animation and sonic output are driven by the same underlying digital-physics parameters, so that participants perceive a single coherent ``force'' rather than separate visual and auditory channels. From an initial pool of brainstormed candidate metaphors (e.g., chains, cloth, bubbles, flows, swarms), we selected three---Rope, Spring, and Magnetic Field---because they are immediately legible from everyday physical experience and each foregrounds a distinct relational dynamic useful for nudging (Table~\ref{tab:mro}). Across all three, mappings are designed to be legible but not fully determined: participants can quickly grasp basic cause--effect relationships while retaining room for exploration and surprise. The OJ may adjust both mapping parameters and mapping relations in real time during performance. Appendix~\ref{sec:mro} provides full specifications for each MRO.

\paragraph{Rope} The Rope is a semi-flexible virtual tether anchored at the chest positions of two or more participants (Figure~\ref{fig:rope}). Its visual geometry sways, bends, and droops dynamically in response to participant movement. The system continuously computes the velocity ($v$) and acceleration ($a$) of the rope's midpoint from the tracked head poses. In the default audio mapping, velocity modulates volume (faster swinging $\rightarrow$ louder) and acceleration modulates pitch (sharper turns $\rightarrow$ higher pitch), evoking the kinesthetic experience of swinging a physical rope: auditory intensity rises with more vigorous collaborative effort. The OJ can adjust the rope's mass and width in real time---increasing mass imparts a heavier, more inertial feel, while reducing it yields a lighter, more responsive dynamic. This design nudges participants toward rhythmic entrainment and coordinated swinging.

\paragraph{Spring} The Spring is rendered as vibrating sine waves with gradient colors, connecting the chest positions of two participants (Figure~\ref{fig:spring}). It stretches and compresses as they move apart or together. The system tracks interpersonal distance ($d$), individual heights ($h$), and height difference. Distance controls pitch (increasing distance raises pitch, simulating tension; decreasing distance lowers it, evoking compression), while height modulates a Low-Frequency Oscillator (LFO) rate: when one participant crouches while another stands, the height difference creates tremolo effects in the sound. Additionally, microphone volume or vocalizations are mapped to spring width, encouraging participants to combine voice---speaking or clapping---with spatial movement. This design nudges participants to negotiate proximity, explore vertical dynamics, and invent sound-making gestures.

\paragraph{Magnetic Field} The Magnetic Field creates an ambient particle field around all participants, with red particles representing positive poles and cyan particles representing negative poles (Figure~\ref{fig:magnetic}). Particles flow, cluster, or disperse based on the magnetic relationships between participants, whose chest positions serve as pole centers. The OJ assigns each participant a polarity (positive or negative) and adjusts field strength; flipping a participant's polarity transforms attraction into repulsion. The system computes pairwise distances ($d$) and field turbulence, which control the ``glitchness'' of the background music: closer proximity produces more fragmented, distorted textures, while dispersal normalizes the sound. With more than two participants, a three-body problem emerges---creating chaotic, unpredictable particle behavior that invites exploratory roaming, orbiting, and clustering rather than a single synchronized action.

\section{Our Study}

\begin{figure*}[ht]
    \centering
    \includegraphics[width=1\linewidth]{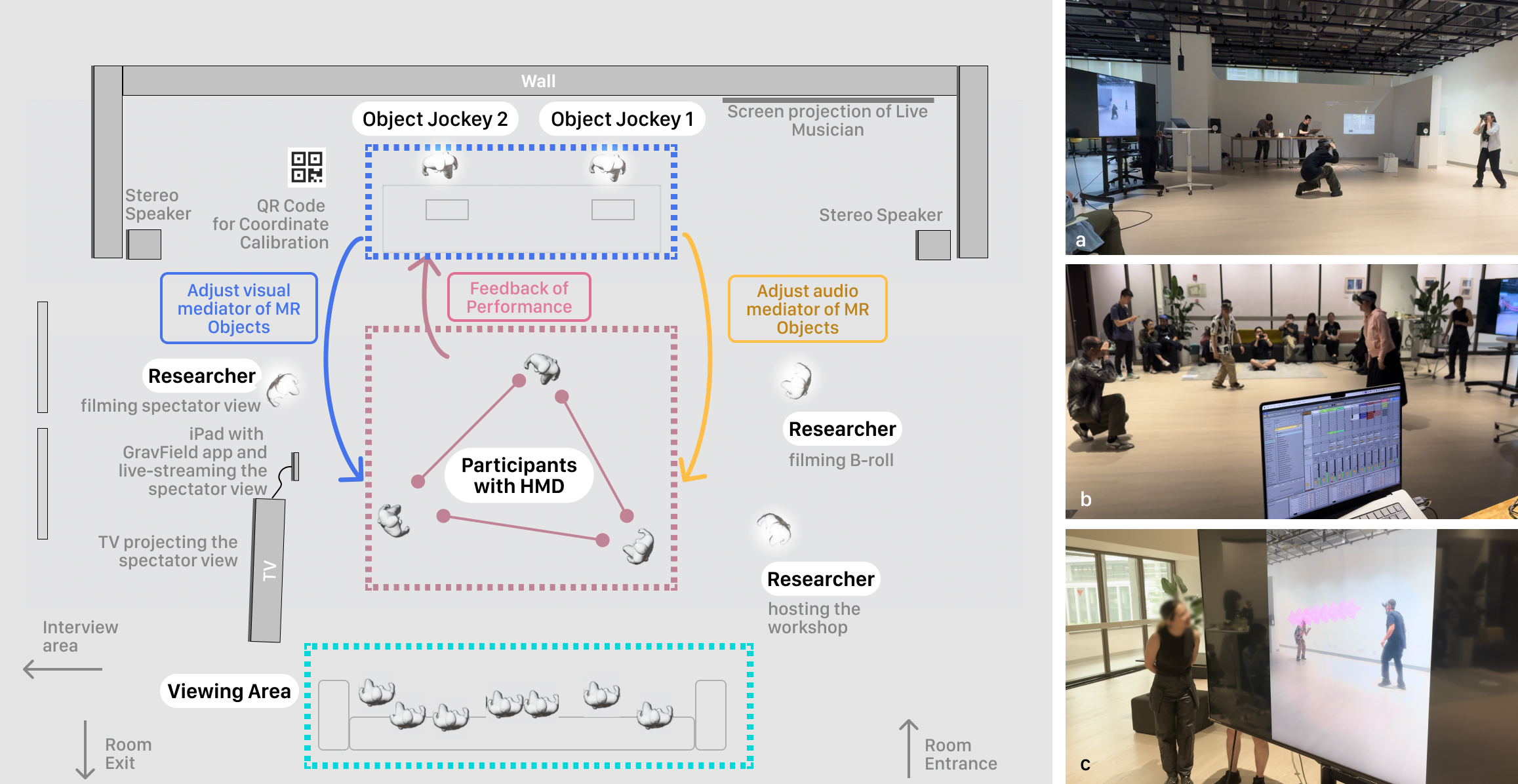}
    \caption{Floor plan of the GravField workshop. OJ workstations sit at the back with stereo speakers on each side; the central area serves as the MR performance stage; a viewing area with TV faces the stage. Colored arrows indicate data flow: visual (blue), audio (orange), and performance feedback (yellow). Insets: (a) audience view of the stage; (b) OJs' view of screens and stage; (c) TV with the live-streamed AR spectator overlay.}
    \label{fig:floorplan}
    \Description{Annotated floor plan of the workshop space (approximately 16 by 8 meters) with three inset photographs. The Object Jockeys' desk with two laptops and audio equipment sits at the top of the plan, with a QR code for spatial calibration nearby and stereo speakers at the far left and right walls. The central open area forms the mixed reality performance stage, and an audience seating area faces the stage at the bottom, with a television showing the iPad spectator view on the left. The three inset photos show: (a) the audience's view of participants on stage, (b) the Object Jockeys' rear view of their screens and the stage, and (c) the television displaying the live-streamed augmented reality overlay.}
\end{figure*}


To investigate how live-configurable MROs nudge collective movement, we adopted a
performance-led, research-through-design (RtD) approach~\cite{Zimmerman2007Research,Benford2013Performance}.
Rather than running a controlled lab experiment, we treated GravField as a
performative research instrument and studied it in use during a two-day
workshop at a digital media conference, facilitated by both Object Jockeys (OJs) together with the research team. This approach allowed us to test the
system with external participants.

\subsection{Participants of the System}

We distinguish clearly between \emph{Object Jockeys (OJs)} and
\emph{workshop participants}, as they play different roles in the ecology of the
system.

\paragraph{Object Jockeys (OJs)}
Two expert practitioners (OJ1 and OJ2; Table~\ref{tab:oj-info}) co-developed the system with the research team and operated it during the workshop. Both have substantial prior experience with live coding, sound art, and MR performance. We treat the OJs as skilled operators of a performance instrument—analogous to DJs using a DJ setup—not as study participants, though we include their behavior in the data. Given GravField's steep learning curve, we decided not to hire new OJs to learn it from the ground up. Our study relies on the OJs' interventions, which may constrain generalization. Their interventions are part of the research apparatus: we analyze how their live configuration shapes the affordance landscape.

\paragraph{Workshop participants}
We recruited 25 volunteers (P1--P25; Table~\ref{tab:participants-info}) from the
conference and the host university's media-arts and HCI communities via a call
posted two weeks in advance. We intentionally targeted people with backgrounds
in movement, music, live coding, or interactive art, because (i) the workshop
relied on comfort with being seen on a stage and improvising in front of many others, and
(ii) such practitioners could articulate contrasts between GravField and their
existing practices. Participants could sign up alone or with friends or colleagues;
thus some groups comprised pre-existing acquaintances while others were
strangers. We treated the \emph{group} as the analytic unit and noted
acquaintance relationships in our memos, returning to them during analysis
when they appeared to influence coordination. We discuss this expert, partially
acquainted sample as a limitation for generalization in Section~\ref{sec:limitation}.

The workshop protocol was approved by the ethics board of one author's
institution. All participants provided informed consent, could withdraw at any
time, and were told there was no ``correct'' way to move: the goal was to
follow what felt interesting or natural in relation to the MROs and to each
other.

\subsection{Workshop Setting and Procedure}

The workshop took place in a large studio classroom (approximately $16\times8$
m) configured as a shared performance space (Figure~\ref{fig:floorplan}). A
central open area served as the mixed-reality ``stage''; the OJs' workstation
was located against the back wall with laptops and audio equipment. Two stereo
speakers provided room-scale sound, and a large TV displayed a live spectator
view (iPad) showing the MRO overlays.

We deliberately staged GravField as a semi-public (open-call) workshop rather than a private
lab session. At any time, one small group (usually two or three participants)
wore MR headsets and performed with a sequence of MROs, while the remaining
participants sat or stood as audience. This \emph{witnessing} served two
purposes: (i) it mirrored participants' existing experience with dance or
music performance, allowing them to compare GravField to familiar stage
contexts, and (ii) it created shared reference points that later enriched
group discussion, as participants could juxtapose their felt experience with
what they had seen others do.

Upon arrival, participants signed consent forms and watched a short
demonstration: two researchers performed with the Rope and Spring MROs, and then a third researcher joined to demonstrate the Magnetic Field. We emphasized that there were
no prescribed moves and invited participants to explore. Participants then
self-organized into eight groups (A--H). Each group typically experienced the
three MROs in sequence---Rope, Spring, Magnetic Field---for a total of around
15--20 minutes per round, with one or two OJs orchestrating in real time.
Because of headset constraints, most rounds involved three participants; one
round involved four.



After each round, a researcher guided participants to a separate interview area and conducted a 15--20 minute post-session interview using (i) video replay of the just-completed session to trigger memory and (ii) soma trajectory sketches to externalize felt qualities of movement and sensation. Interviews used semi-structured prompts (Appendix~\ref{sec:questions}) and encouraged collaborative reflection on first-person experience and social negotiation. After all groups completed the experience, we held a roughly 50-minute reflective group discussion with all participants and both OJs; in this discussion, OJs articulated their intentions and decision-making during live configurations, allowing us to examine nudging as an explicit, situated practice rather than an opaque system process.

\begin{table*}[ht]
\centering
\begin{tabular}{@{}lllll@{}}
\toprule
\textbf{Group} & \textbf{ID} & \textbf{Gender} & \textbf{Age Range} & \textbf{Background} \\
\midrule
A & P1  & Male   & 20-30 & Musician \\
& P2  & Male   & 30-40 & New Media Artist \\
& P3  & Male   & 30-40 & Professor in New Media Art \\
\hline B & P4  & Female & 30-40 & Dancer and Educator \\
& P5  & Male   & 20-30 & Professional Live Coder \\
& P6  & Male   & 30-40 & Dancer \\
\hline C & P7  & Female & 20-30 & Graduate Student in HCI \\
& P8  & Female & 20-30 & Graduate Student in Media Art \\
& P9  & Male   & 30-40 & New Media Artist \\
\hline D & P10 & Male   & 20-30 & Graduate Student in Music \\
& P11 & Male   & 20-30 & Entry-Level Live Coder \\
& P12 & Female & 20-30 & Graduate Student in New Media Art \\
\hline E & P13 & Female & 20-30 & Undergraduate Student in HCI \\
& P14 & Female & 20-30 & Undergraduate Student in HCI \\
& P15 & Male   & 30-40 & New Media Artist and Lecturer \\
\hline F & P16 & Female & 30-40 & New Media Artist and Lecturer \\
& P17 & Female & 30-40 & Professor in HCI \\
& P18 & Male   & 20-30 & Teaching Assistant and Artist in New Media Art \\
\hline G & P19 & Female & 20-30 & Live Coder \\
& P20 & Female & 20-30 & Media Artist \\
& P21 & Female & 20-30 & Experimental Musician/Sound Therapist \\
\hline H & P22 & Male   & 20-30 & New Media Artist \\
& P23 & Male   & 20-30 & New Media Artist \\
& P24 & Female & 20-30 & Researcher in HCI \\
& P25 & Female & 20-30 & Sound Artist \\
\bottomrule
\end{tabular}
\caption{Participants' Information}
\label{tab:participants-info}
\Description{Table listing 25 workshop participants across five columns: Group (A--H), Participant ID (P1--P25), Gender, Age Range, and Background. The group includes 12 male and 13 female participants, all aged 20--40. Backgrounds span live coding, new media art, dance, sound design, music, and interaction design.}
\end{table*}

\begin{table*}[ht]
\centering
\begin{tabular}{@{}lllll@{}}
\toprule
\textbf{Group} & \textbf{ID} & \textbf{Gender} & \textbf{Age Range} & \textbf{Background} \\
\midrule
OJ 
& OJ1 & Male & 30-40 & Sound Artist, Live Coder \\
& OJ2 & Male & 30-40 & New Media Artist, MR Engineer \\
\bottomrule
\end{tabular}
\caption{Object Jockeys' Information}
\label{tab:oj-info}
\Description{Table listing the two Object Jockeys across four columns: Role, Gender, Age Range, and Background. Object Jockey 1 is male, aged 30--40, with backgrounds in new media art and music. Object Jockey 2 is male, aged 30--40, with backgrounds in new media art and mixed reality engineering.}
\end{table*}

\subsection{Data Collection}

We collected a multi-perspective dataset designed to capture bodily, sonic, and
social dynamics:

\begin{itemize}
  \item \textbf{Spectator and ambient video:} Wide-angle videos and the
  MR spectator view (iPad) recorded participants' movements and the MRO
  overlays. Additional B-roll cameras focused on the OJs' workstation
  (Unity, TouchOSC, Ableton), capturing when and how they intervened.

  \item \textbf{Audio:} We recorded both the system's audio output and
  room sound, including participants' utterances and vocalizations.

  \item \textbf{Post-session interviews:} After each round, we conducted
  15--20 minute in-situ group interviews, using immediate video replay and
  soma-trajectory drawing prompts~\cite{Tennent2021Articulating} to help
  participants reconstruct the temporal arc of their experience.

  \item \textbf{Final group discussion:} At the end of the workshop, we
  ran a 50-minute plenary discussion in which participants and OJs
  reflected on emergent patterns, roles, and the experience of live
  orchestration.

  \item \textbf{Field notes and analytic memos:} Researchers documented
  observations after each session, including perceived energy shifts,
  notable OJ interventions, and relational configurations (e.g., friends
  vs strangers).
\end{itemize}

Across the two days we collected 6.5 hours of interaction video, 1.5 hours of
small-group interviews, 2 hours of large-group discussion, 8 sets of
soma-trajectory drawings, and associated field notes.

\subsection{Data Analysis}

We employed a bricolage analysis strategy~\cite{Rogers2015Contextualizing}, combining
video, audio, drawings, transcripts, and OJ reflections. Our interdisciplinary
team (HCI researchers, MR designers, a game designer, and the two OJs) first
reviewed the material to identify salient episodes of coordination, breakdown,
and re-engagement.

We began with open coding of interview transcripts and group discussion to
identify recurring themes (e.g., ``probing the object'', ``falling into
shared pulse'', ``getting tired'', ``waiting for something new''). In
parallel, we annotated videos with visible indicators of OJ interventions,
including (i) OJ--participant talk, (ii) changes in MRO behavior on screen,
(iii) abrupt shifts in the soundscape, and (iv) OJ activity visible in
B-roll. We used these cues to approximate when OJs reconfigured the system,
and related them to shifts in participants' coordination and reported
experience. Because we did not log parameters in real time, our analysis
focuses on \emph{patterns of intervention} rather than precise counts.

We then reconstructed soma trajectories for each group by aligning their
drawings with the annotated videos, focusing on changes in perceived
interestingness, familiarity, inclusion of other in the self, and sense of
agency over time. Finally, we iteratively refined our themes in relation to
the research questions (RQ1 and RQ2), treating groups---rather than
individuals---as the unit of analysis. Throughout, we triangulated
participant accounts with observable behavior and OJ autoethnographic notes,
and we conducted a member-check with both OJs to validate our interpretation
of the live orchestration practice.

\section{Results}

\begin{figure*}[ht]
  \centering
  \includegraphics[width=0.7\linewidth]{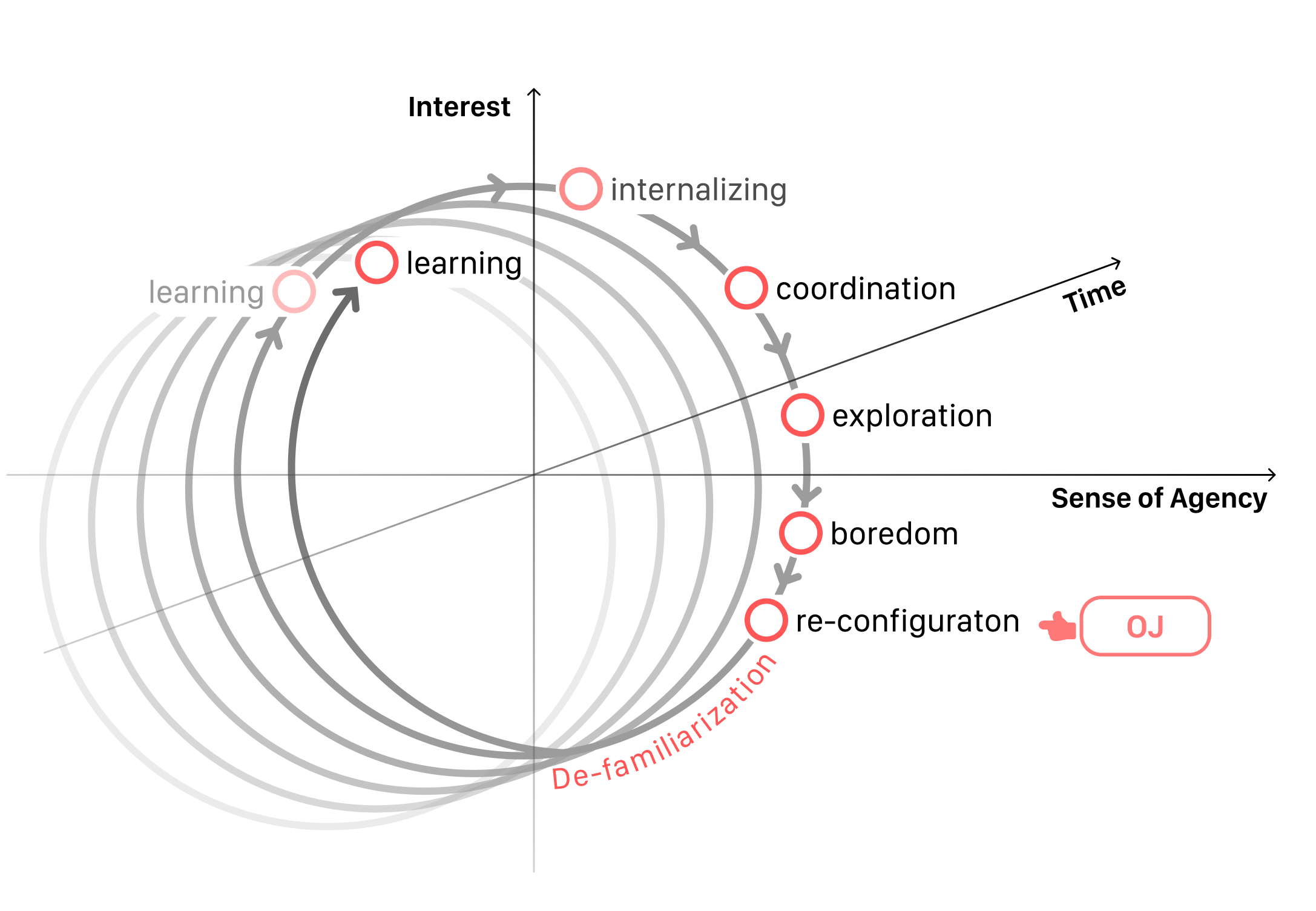}
  \caption{Live Nudging Spiral. Participants progress through six stages---learning, internalizing, coordination, exploration, boredom, and reconfiguration---mapped along two axes: interest (vertical) and locus of control (horizontal). When engagement plateaus at boredom, the OJ live-configures the MRO to reignite exploration, restarting the cycle.}
  \label{fig:spiral}
  \Description{Circular diagram illustrating the six-stage Live Nudging Spiral. Six labeled nodes are arranged clockwise around a circle: Stage 1 Learning at the top-left, Stage 2 Internalizing at the top-right, Stage 3 Coordination at the right, Stage 4 Exploration at the bottom-right, Stage 5 Boredom at the bottom-left, and Stage 6 Re-configuration at the left. Curved arrows connect each stage to the next in clockwise order, forming a continuous cycle. Two axes cross the center of the circle: a vertical axis labeled Interest with low at the bottom and high at the top, and a horizontal axis labeled Control with participant control on the left and Object Jockey control on the right. Learning and Internalizing sit in the high-interest, participant-control quadrant; Coordination and Exploration sit in the high-interest, Object-Jockey-control quadrant; Boredom sits in the low-interest region. A highlighted arrow from Boredom to Re-configuration marks the Object Jockey's live intervention, and a return arrow from Re-configuration loops back to Learning, restarting the cycle.}
\end{figure*}

\begin{figure*}[ht]
  \centering
  \includegraphics[width=0.67\linewidth]{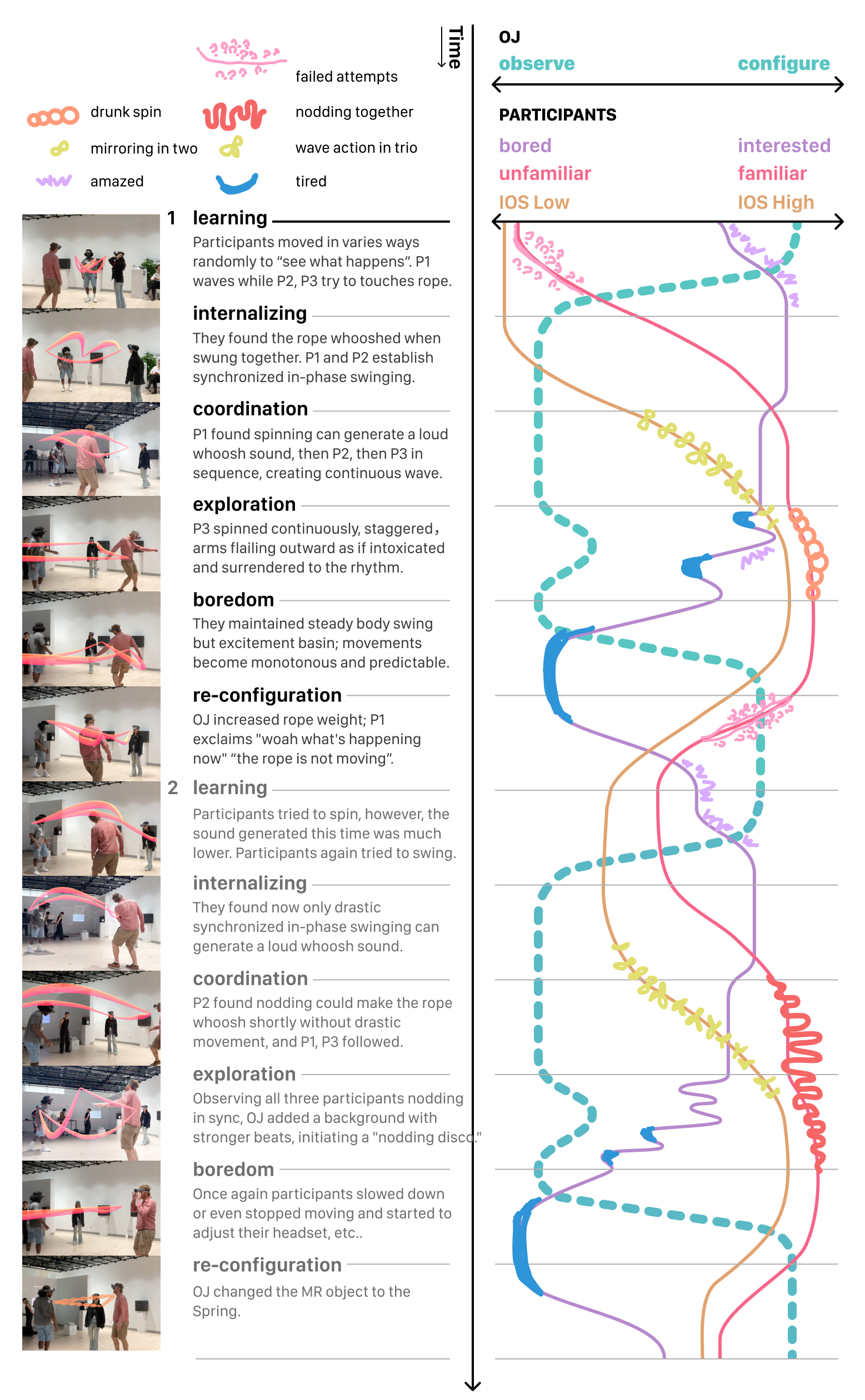}
  \caption{Soma Trajectory of a GravField session. Colored curves trace how perceived interestingness, familiarity, and Inclusion of Other in the Self (IOS) evolve over time, reconstructed from participants' post-session reflections. Annotations mark key embodied behaviors at each stage of the Live Nudging Spiral; the dashed curve shows the OJ's balance between observing and configuring.}
  \label{fig:soma_trajectory}
  \Description{Time-series plot with time on the horizontal axis and four colored curves. A blue curve for perceived interestingness rises during learning and exploration then dips at boredom. A green curve for familiarity climbs steadily before plateauing. An orange curve for Inclusion of Other in the Self peaks during coordination. A dashed gray curve for the Object Jockey's observe-versus-control balance spikes at re-configuration moments. Vertical dashed lines divide the plot into the six Live Nudging Spiral stages, with text annotations marking representative behaviors such as tentative probing, synchronized swinging, repetitive slowing, and parameter change.}
\end{figure*}

We found that participants' performances unfolded as an evolving cycle of discovery, coordination, and rediscovery, which we organized into a \emph{``Live Nudging Spiral''}: periods of increasing group attunement and interest eventually gave way to saturation or boredom, at which point the OJs reconfigured MROs to renew participants' curiosity. We organized our qualitative findings around two analytic foci: (RQ1) how MROs nudged intercorporeal movement and shaped collective behaviors; and (RQ2) how OJs’ live reconfiguration of MROs sustained open-ended, exploratory participation over time.

\subsection{Temporal dynamics: Live Nudging Spiral}

Soma trajectory reconstructions charted participants' perceived \emph{interestingness}~\cite{Silvia2005What}, \emph{familiarity}~\cite{Whittlesea1993Illusions}, \emph{inclusion of other in the self} (IOS)~\cite{Aron1992Inclusion}, and \emph{sense of agency}~\cite{Tapal2017Sense}, alongside OJs' \emph{observe/control} over time. This helped us locate turning points in performance sessions. Figure~\ref{fig:soma_trajectory} shows an excerpt of a typical session with Rope MROs. By analyzing participants' post-hoc \emph{``soma trajectories''} and video recordings across sessions, we identified a recurrent six-stage arc that we term the \emph{Live Nudging Spiral} (see Figure~\ref{fig:spiral}):

\paragraph{(S1) \textbf{Learning}---Probing to sense digital physics and the intentionality of nudging.}  When participants first encountered an MRO, they were unfamiliar with the \emph{``digital physics''} behind it. They used tentative, playful movements to probe and sense response characteristics. Through this process, participants began to understand not only the affordance of the MRO (how it can be interacted with) but also the intentionality of the OJs' nudging (what kinds of interaction are more \emph{``rewarded''}). The audiovisual feedback scaffolded rapid hypothesis-testing of these unfamiliar dynamics (see~\S\ref{sec:rq1_unfamiliarity}--\ref{sec:rq1_affordances}).

\paragraph{(S2) \textbf{Internalizing}---Perceptual coupling and embodied mastery.}

In this stage, participants transitioned from unfamiliar to familiar by internalizing the digital physics they had learned through perceptual coupling. Through repeated movements and feedback, as action--feedback contingencies stabilized, participants began treating the MRO's dynamics instinctively as part of their body schema and developed embodied mastery of interaction patterns. Participants reported perceiving force through sound, which led to more economical strategies as they discovered the ``sweet spots'' of each MRO (see~\S\ref{sec:rq1_feedback}--\ref{sec:rq1_mastery}).

\paragraph{(S3) \textbf{Coordination}---Group negotiation and social entrainment.}

Once perceptual coupling was in place, groups began to negotiate agency, and coordination emerged through social entrainment. Participants used MROs as a medium for negotiation; some dyads discovered that synchronizing their movements would amplify the MRO's response, aligning with the OJs' nudging intention. This resulted in emergent dyadic coordination. For trios, we observed \emph{relay} patterns as participants created sequential flows of movement. These behaviors were emergent, not instructed (see Figure~\ref{fig:groupsize}; \S\ref{sec:rq1_coordination}).

\paragraph{(S4) \textbf{Exploration}---Expressive divergence.}

Once stable coordination was established, participants deliberately perturbed it to widen the expressive space. When participants discovered that their actions could alter the MRO's audiovisual output, they experimented with unexpected inputs---vocalizations, exaggerated gestures, and novel movement styles---to observe how these affected the MRO's appearance and sound. This controlled destabilization expanded the shared repertoire while preserving a negotiable center of rhythm (see~\S\ref{sec:rq2_ambiguity}--\ref{sec:rq2_creative}).

\paragraph{(S5) \textbf{Boredom}---Fatigue, cognitive saturation, and interestingness plateau.}

As novelty decayed or effort accumulated, variability collapsed into metronomic repetition and engagement plateaued. At this stage, participants had exhausted the pattern; they no longer felt interested or engaged due to cognitive saturation or physical fatigue as they depleted their search space or energy with a given configuration. Some participants repeated their movements at a slower pace, while others disengaged and began adjusting their headsets. The OJs could sense this shift and felt urged to intervene (see~\S\ref{sec:rq2_boredom}).

\paragraph{(S6) \textbf{Re-configuration}---Defamiliarization and relearning.}

After the boredom stage, the OJs usually perturbed the affordance landscape (e.g., remapping, parameter shifts, sound effect changes) to defamiliarize~\cite{shklovsky1917art} participants' perceptual coupling, reopen the search space, and catalyze a fresh cycle of the spiral, thereby sustaining open-ended exploration. With embodied mastery, participants could quickly sense subtle nudging and reacted to the changes. For more drastic reconfigurations, such as reverse-remapping sensory feedback to discourage previously rewarded movements, participants had to relearn and adapt to the new affordance landscape. The cycle then spiraled back to the learning stage with renewed perceptual coupling (see~\S\ref{sec:rq2_reconfig}).

\begin{figure*}[ht]
  \centering
  \includegraphics[width=1\linewidth]{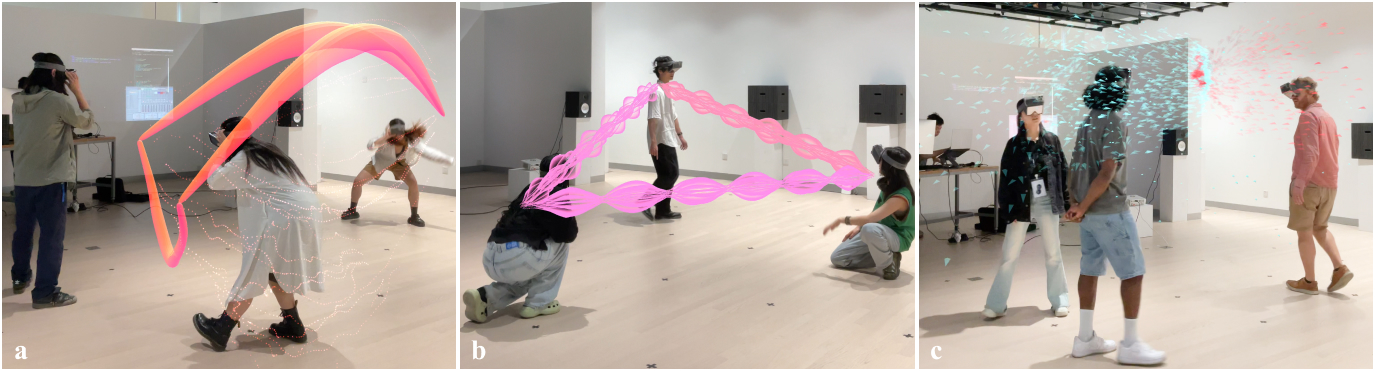}
  \caption{Characteristic movement patterns elicited by each MRO: (a) the Rope encourages dynamic swinging and mirrored coordination; (b) the Spring heightens awareness of vertical displacement and push--pull exchanges; (c) the Magnetic Field prompts fluid, exploratory navigation.}
  \label{fig:different}
  \Description{Three-column photo grid comparing participant movement behaviors across the three mixed reality objects. Column (a), Rope: four images of participants making large swinging and spinning movements, with a glowing pink rope arcing widely between them. Column (b), Spring: four images of participants squatting and rising, with a taut pink ribbon stretching vertically to emphasize height-based interaction. Column (c), Magnetic Field: four images of participants making slower, exploratory movements through scattered blue and red floating particles.}
\end{figure*}

\subsection{RQ1: How Mixed Reality Objects Nudge Collective Movement}

\subsubsection{Unfamiliarity as Entry Point}
\label{sec:rq1_unfamiliarity}
When first encountering a virtual object in the performance, participants often experienced a brief period of hesitation and disorientation. The MROs had no explicit instructions, so initial interactions were driven by curiosity and basic sensorimotor inquiry. Participants approached the MRO cautiously at first—reaching out toward a floating rope end or testing the “spring” connecting them to a partner. This unfamiliarity served as an entry point into the experience: unsure of what the object could do or how it would respond, users engaged in exploratory touching and tentative movements to “feel out” the object’s properties. For example, with the Rope MRO, several participants gently tugged or swung the virtual rope in small motions to gauge its behavior. Some wondered aloud whether the rope would \emph{``resist or pull [them] back''} if they tugged, indicating an initial uncertainty about the rope’s dynamics. Such moments of not knowing what would happen next heightened participants’ bodily awareness and primed them for discovery.

\subsubsection{Intuitive Physical Affordances}
\label{sec:rq1_affordances}
Despite the novelty of the medium, participants quickly picked up on the physical metaphors suggested by each virtual object’s appearance and behavior. The design of the objects leveraged familiar real-world affordances, and indeed we observed participants naturally acting on those cues. The Rope, for instance, visually dangled with a certain heft in augmented space, immediately inviting a swinging action. As P4 remarked, \emph{``It [was] hanging down like a real rope with weight---it just looked like something you’d want to swing.''} True to this intuition, participants instinctively grabbed at their end of the Rope and began oscillating it, as if initiating a playground game of jump-rope. Likewise, the Spring’s visual form (a taut line connecting two people) prompted users to bounce and bob in place, exploring tension and release. Even without any explicit prompt, people treated the virtual spring as if it would stretch and recoil, squatting down experimentally or giving quick upward hops. These immediate, embodied responses illustrate how the perceived materiality of the MR objects nudged people’s movements in expected directions. The objects’ affordances acted as a tacit guide: participants read the Rope, Spring, and Magnetic Field as analogous to their physical counterparts (a heavy rope, a springy tether, a field of pull), and this reading shaped their first attempts at interaction.

\subsubsection{Synesthetic Feedback Loops Reinforcing Action}
\label{sec:rq1_feedback}
Once participants began manipulating an MRO, the system’s multisensory feedback quickly created a self-reinforcing loop between their movements and perceptions. Each object produced responsive audiovisual effects that were tightly coupled to user actions, aligning with natural physics. This “synesthetic” mapping---where exertion could be heard and seen---encouraged further exploration by making the consequences of movement viscerally felt. Participants swinging the Rope soon noticed that its motion had an accompanying sound: a low “whoosh” that intensified in volume and pitch the faster they moved. This feedback delighted users and clarified the rope’s dynamics. As P3 noted, \emph{``The sound feedback was very obvious as I swung the rope. It was really rewarding. I think I should swing faster.’’} \emph{``I could hear how fast I was moving the rope,’’} said another participant (P7), emphasizing that the rope’s resistance was not only seen but also audibly perceived. Another described it as \emph{``hearing the force''} of their swing (P21). Similarly, with the Spring MRO, participants discovered that changing their relative height or distance altered an emitted tone---stretching the spring by stepping back made the sound swell, while a sudden drop into a crouch produced a downward slide in pitch. These real-time sensory responses matched participants’ intuitive expectations (e.g., a tighter spring yielding a higher-pitched strain sound), which in turn validated and amplified their bodily experimentation. The more energetically they moved, the richer the audiovisual payoff, leading them to further push the limits of the object. Through these tightly coupled feedback loops, the virtual objects effectively “communicated” their physics to participants, teaching them how to engage the object through the body.

\subsubsection{Familiarization and Embodied Mastery}
\label{sec:rq1_mastery}
As the sessions progressed, participants moved from tentative probing to more confident, embodied mastery of the objects’ interaction patterns. With repeated action and feedback, they internalized how their movements influenced the MRO and began to optimize for more pronounced effects. In the Rope sessions, for example, participants initially experimented with gentle back-and-forth swings. But after hearing only subtle sounds from small motions, they grew bolder. Some discovered that more dramatic motion---especially spinning their body around to whip the rope in a circle---generated a far louder and more exciting “twang.” One participant reflected on this learning process: \emph{``When I first tried swinging it back and forth, the sound was subtle, but [doing a] 360° spin made it much louder''} (P18). Through such trial and error, participants learned the “sweet spots” of interaction that maximized the object’s response. Interestingly, these emergent techniques were not explicitly instructed by the system; rather, the virtual object’s properties nudged users toward them. For instance, the Rope’s weight and sound model rewarded faster, continuous spinning---an action the designers had not specifically requested, but which the participants organically adopted once they realized it yielded a satisfying result. In this way, familiarization with the MRO’s dynamics led to increasingly fluent and inventive movements. Participants began to treat the virtual objects almost like extensions of their own bodies, anticipating their behavior and skillfully manipulating them to achieve desired sensory outcomes.

\subsubsection{Object-Specific Patterns of Movement}
Each type of MRO catalyzed a distinctive pattern of group movement, highlighting how different digital “physics” can shape social choreography (Figure~\ref{fig:different}). With the Rope MRO, two participants found themselves effectively engaged in a cooperative tug-and-swing dynamic. Initially, they might move asynchronously, but very soon a mirrored rhythm emerged: both partners synchronizing their swings to make the rope arc higher and generate louder swooshes. We observed pairs gradually fall into phase with each other, pulling in unison as if jointly pumping a playground swing. This mirroring behavior was encouraged by the rope’s design, which made momentum cumulative---one person’s force contributed to the overall motion, inviting the other to match it for greater effect. In contrast, the Spring MRO introduced an oppositional yet complementary movement pattern. Participants connected by a spring often assumed facing stances and took turns bending their knees or backing away, effectively creating a push--pull seesaw of motion. When one participant squatted down, the other would seize the moment of slack to approach or rise up, and vice versa; this alternating dance produced a tense interplay, accompanied by the spring’s changing audio feedback that signaled strain. The spring heightened awareness of both vertical and horizontal distance between partners, leading to a playful “confrontation”. Meanwhile, the Magnetic Field MRO---a cloud of virtual particles that orbited around participants---prompted more fluid and exploratory movements rather than tight synchrony. Lacking a single obvious physical analogy, the magnetic field’s influence was ambiguous, and participants responded by roaming curiously around the space. They watched the floating particles trail behind themselves and others, experimenting with how coming close or moving apart affected the swarm. A common tendency was to walk in circular or orbital paths, as if tracing magnetic lines of force. \emph{``I felt like an electron orbiting… I would drift towards one person then get repelled to orbit another,''} one participant (P5) vividly described. Notably, the Magnetic Field did not nudge the group into one cohesive motion; instead, it created an ambient collective awareness---participants were moving in response to the field and its audio feedback as much as (or more than) directly mirroring each other. In summary, each MRO shaped a characteristic form of interaction: Rope encouraged mirrored, energetically coupled movement; Spring engendered oppositional, tension-and-release exchanges; and the Magnetic Field invited open-ended, interpretive movement and observation (Figure~\ref{fig:different}a--c).

\subsubsection{Emergent Group Coordination}
\label{sec:rq1_coordination}

\begin{figure*}[ht]
  \centering
  \includegraphics[width=0.58\linewidth]{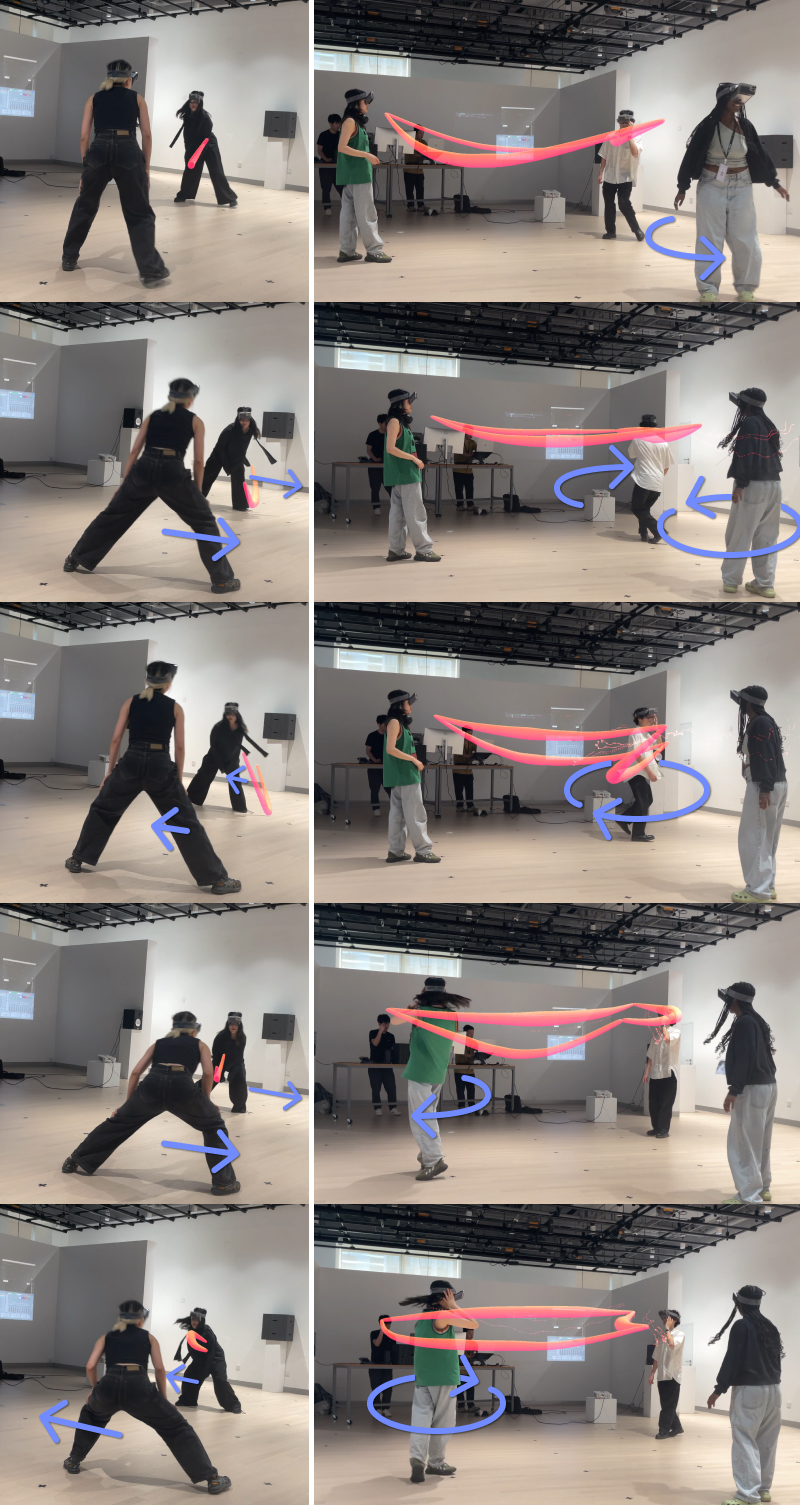}
  \caption{Effect of group size on coordination patterns with the Rope MRO. Left: paired participants tend to mirror each other's movements symmetrically; right: trios often develop a relay-like sequence in which one person initiates a spin, the next adopts it, and the third follows.}
  \label{fig:groupsize}
  \Description{Two-column photo grid comparing dyadic and triadic interactions with the Rope mixed reality object. Left column (pairs): three sequential photos of two participants facing each other and performing synchronized mirroring swings, with the virtual rope arcing symmetrically between them. Right column (trios): three sequential photos of three participants in a triangle formation performing a relay pattern, where each person in turn initiates a spinning motion that passes sequentially around the group.}
\end{figure*}

When all participants engaged with the MROs, we observed new coordination patterns emerging after constant negotiation of agency. All participants quickly found ways to be involved in the joint action. For example, in the Rope MRO session, one participant would begin nodding to the rhythm, which effectively made the rope swirl while requiring minimal effort. Other participants gradually adopted the same motion. Initially, they moved at different paces, producing fractured, low sounds. Over time, they established a shared pulse through subtle head nods and gentle body sways that gradually synchronized. The rope then achieved wider swings and produced louder sounds on beats. P8 described this experience: \emph{``It felt like we were all tuning into the same invisible metronome.''} This coordination typically emerged after 45--60 seconds of interaction, suggesting participants needed time to attune to each other's movements. Another example is the relay pattern: More prominently in longer sessions (3+ minutes), participants created a sequential flow of movement around their circle. For example, with the Rope, one participant would initiate a spin, creating distinct audiovisual feedback. The closest participant would then adopt this movement, followed by the third person, creating a visible wave through the group (A→B→C→A; see Figure~\ref{fig:groupsize}). During this pattern, non-active participants would observe quietly, allowing each person to experience their moment of influence without interruption. Remarkably, the timing between transitions became consistently regular (estimated from video review at approximately 1--2 seconds) despite no explicit instructions to maintain such timing. This relay pattern also manifested in the Spring configuration, where the wave took the form of sequential squats or claps (A squats→B squats→C squats→A squats again). These higher-order group patterns were not explicitly built into the system's script but arose organically from participants' mutual adaptation. The desire to include everyone spurred participants to improvise novel group behaviors (like the spinning relay or a three-person pulsing wave), demonstrating how the MRO's subtle nudges scaled to support whole-group movement. In sum, the mixed-reality objects provided just enough structure to spark coordinated activity while leaving space for users to co-create complex group choreography on the fly.

\subsection{RQ2: How Live Configuration Sustains Open-Ended Exploration}

\subsubsection{Non-Coercive Ambiguity Encouraging Exploration}
\label{sec:rq2_ambiguity}
A key aspect of GravField’s design is that it nudges without prescribing, creating a space where participants feel at liberty to play. Indeed, once participants moved beyond the initial learning curve, their behavior often became highly creative and idiosyncratic. The absence of explicit “correct” moves meant participants felt no fear of “wrong” interactions---a psychological safety zone in which people felt free to experiment with their whole bodies and even voices. The system’s deliberate ambiguity in how feedback was provided further ensured that users did not settle into any one routine too quickly. Instead, the open-ended nature of the MROs led different individuals to explore the same digital parameters in markedly different ways. For instance, participants working with the Spring discovered an unconventional method of interaction: they realized that the virtual spring’s visual amplitude could be influenced not only by physical movement but also by sound. In one session, noticing that loud noises made the spring’s oscillation visibly bigger, participants began vocalizing---calling \emph{``hellooo,''} imitating drum beats, and even issuing comical shouts---to see how their voices might “shake” the spring. Rather than restricting themselves to bodily motions, they co-opted the environment’s audio sensitivity as another channel for play. Another participant (P6), while wielding the Rope MRO, experimented with varying his movement style to push the limits of the rope’s feedback. He transitioned from straightforward swings to exaggerated, almost theatrical motions---at one point staggering and swaying like a drunk person for dramatic effect---simply to observe how the rope’s sound responded. From the OJs’ perspective, \emph{``ambiguity supported playful countermoves’’} (OJ1); this controlled destabilization expanded the shared repertoire while preserving a negotiable center of rhythm. In short, because the system did not dictate a single way to interact, participants felt empowered to probe the boundaries of what was possible, using imagination and even humor in their exploratory actions. The result was a rich diversity of behaviors emerging from the same basic set-up, driven by each participant’s personal curiosities and intuitions.

Interestingly, we found less visual guidance enhanced creativity. In sessions with P19--P25, OJs temporarily disabled visuals, using only sound feedback.

Participants consistently noted that visual feedback actually limited their imaginative potential compared to the audio-only experience. With P22--P25, we observed a powerful moment during a Spring session without visuals: P22 lay on the floor, lowering the sound pitches, while others moved in circles around him; and he reached out with a pleading gesture toward the others as if asking them to return. This emotionally rich interaction—which participants later described as \emph{``a heartbreak moment of separation''}—became impossible once the Spring visual was introduced, as the explicit connection constrained their interpretive creativity and foreclosed certain embodied metaphorical expressions. As P24 reflected, \emph{``Without seeing the spring, I could imagine it as anything—a connection, a relationship, a feeling, or the absence of the connection. But once I saw it, it was just...a spring.''} 

While visual feedback provides initial guidance, it restricts creative possibilities that emerge from ambiguity. The abstract soundscape \emph{``left more to the imagination,''} encouraging users to explore unseen dynamics through novel bodily expressions.

\subsubsection{Creative Expressions and Emergent Social Play}
\label{sec:rq2_creative}
The freedom afforded by live-configurable MROs also enabled spontaneous social and expressive acts that went beyond what the designers initially envisioned. Participants not only manipulated the objects in creative ways, but also invented new shared practices and gestures on the fly. 

In one Magnetic Field session, two friends (P20, P21) found themselves drawn together by the virtual “magnet” forces swirling around them; when they drifted physically close, they responded playfully by giving each other a quick kiss on the cheek, as if the technology had orchestrated a moment of affection. In another instance, two college students (P13, P14) discovered they could collectively perform an orbital dance in the Magnetic Field: standing face-to-face, they joined all four of their hands together and began spinning as a duo at the center of the field. This impromptu waltz was entirely self-initiated---an appropriation of the system to create a new movement pattern (a human “planetary orbit”) not suggested by any interface prompt. As one participant later explained of that moment, \emph{``We felt like we were planets rotating around each other---it just made sense in that moment.''} This quote underscores how the ambiguous nature of the MR environment invited users to attach their own meaning and narrative to the experience. 

Similarly, participants often established informal game “rules” or roles during exploration. With the Spring MRO, a common convention emerged where one person would hold still to let the other experiment freely for a time. However, some deviated from this pattern: one participant (P10) chose to move in the opposite direction while mirroring his partner’s vertical bouncing. When interviewed about this unorthodox tactic, he said he was trying to \emph{``counterbalance [the] parameter changes so that the sound could maintain an optimal tonal quality.''} In effect, P10 treated the interaction like a collaborative improvisation, actively managing the audiovisual output to achieve a pleasing result. Such anecdotes illustrate participants’ capacity to co-create on top of the given system---inventing aesthetic flourishes, cooperative timings, and even intimate interactions, all of which sustained their engagement by keeping the experience fresh and personally resonant.

\subsubsection{Negotiating Agency with OJ}
Open-ended exploration sometimes led participants to test the boundaries of who (or what) was in control of the experience. We observed people playfully negotiating agency with the system and the human facilitator, turning the performance into a two-way conversation. For example, one participant (P14) in a Magnetic Field session began dramatically “conducting” the floating particles with sweeping hand gestures---as if she could marshal the swarm by will---even though the system did not explicitly map any controls to those movements. The OJs, noticing this emergent performance, chose to support it: OJ1 subtly adjusted the sound parameters in real time so that the auditory effects coincidentally responded to the participant’s improvised hand motions. Later, OJ1 described this intervention as \emph{``a bit of wizardry in service of participant creativity,''} essentially hacking his own system on the fly to give the user a momentary sense of magisterial power. This highlights the fluid power dynamic in the live setup: participants were not merely being led by the technology or the OJs, but could also influence the unfolding interaction through their imaginative actions---with the OJ ready to adapt and follow their lead. In interviews, OJs noted that they saw themselves as co-performers rather than strict directors, listening for participants’ initiatives and then amplifying them. This flexible, improvisational relationship between designer and user helped sustain a feeling of shared agency. Participants realized they could bend the experience in unanticipated directions, and that the system (and its orchestrator) would respond in kind.

\subsubsection{Boredom and Fatigue as Signals for OJ Intervention}
\label{sec:rq2_boredom}
Although the open-ended design invited continual exploration, we noted that participants’ engagement could wane after a period of sustained activity. Physical fatigue and perceptual saturation (“boredom”) naturally set in as people exhausted their ideas or energy with a given configuration. For instance, after several minutes of enthusiastic rope swinging, one pair of participants eventually fell into a comfortable, repetitive pattern and then gradually slowed down. \emph{``I felt like I’d tried everything already. I just didn’t know what else to do...''} admitted P5 about this moment of stagnation. Both partners (P4 and P5) had by then achieved a perfectly synchronized back-and-forth swing; as their interest dwindled, their movements became smaller and less coordinated, and they began to disengage from the task. In another session, the limitation was physical endurance: while exploring the Spring MRO’s possibilities, P15 performed a series of exuberant frog jumps to push the spring’s sound feedback to its limits, but soon had to stop, laughing and exclaiming, \emph{``I am exhausted!''} Such episodes---where a once-stimulating activity leveled off into monotony or tiredness---were crucial turning points in each session. They were immediately apparent to the facilitator: the OJs could see when participants stopped proposing new actions or when their body language shifted toward disengagement. OJ1 described, \emph{``I can sense the boredom of the stage; it urges me to do something to make them engaged.''} In these moments, the OJs took on the role of an attentive caretaker of the experience. As the OJ1 put it, \emph{``I felt like I was holding a string connected to everyone in the space. When I saw someone disengaging, I felt obligated to use the [MR] feedback to pull them back in.''} In other words, lulls in participant activity acted as clear signals for the OJs to step in and refresh the interaction, illustrating the collaborative interplay between participant energy and facilitator response in sustaining open-ended engagement.

\subsubsection{Re-Defamiliarization through Live Reconfiguration}
\label{sec:rq2_reconfig}
To reinvigorate the experience at these junctures, the OJs employed real-time reconfiguration of the virtual environment---essentially making the familiar strange again so that exploration could restart. This process often involved tweaking the parameters of the active MRO or layering in new sensory elements, all without pausing the session. The OJs might, for example, increase the background music tempo, introduce a different audio filter, or adjust the \emph{``digital physics''} of an object (such as its weight, elasticity, or force settings); on occasion, the OJs would even swap one MRO for another entirely. These live changes were typically subtle yet unmistakable to participants, who almost invariably responded with renewed curiosity and movement. As OJ1 explained regarding the Rope MRO, \emph{``I adjust the rope's weight and lift the tempo, which expanded participants' steps as they react at a quicker and quicker pace.''}

Beyond adjusting parameters like rope weight mentioned earlier, the OJs could implement a variety of additional changes. In another example, two participants (P7 and P8) interacting with the Spring MRO settled into a small-amplitude bouncing see-saw pattern. When they seemed to be quite accustomed and slowing down in motion, OJ1 subsequently increased the spring's sensitivity to distance, removed the Low-Frequency Oscillator (LFO) rate filter from their average height, and re-linked it exclusively to one participant's height. This perturbation disrupted their previously stable bouncing sensation—for P7, horizontal movement feedback became more pronounced while height changes produced no feedback; for P8, she discovered her squatting movements triggered louder, more varied sound feedback. This led them to experiment with different movements. After a moment of confusion, they formed a new pattern of interaction, with P8 squatting in the center and P7 circling around her, alternating between coming closer and moving farther away.

By altering the feedback loop---be it through sound, visuals, or introducing a novel object---OJ1 effectively re-defamiliarized the scenario. Participants now had something new to learn and play with again, and they eagerly re-entered an exploratory mindset. This tactic of live reconfiguration was used judiciously: OJs waited for clear signs of dwindling engagement before intervening, so as not to disrupt active exploration prematurely. When deployed, however, it proved key to sustaining the open-ended nature of the session. Each reconfiguration kicked off a fresh cycle in the Live Nudging Spiral (Figure~\ref{fig:spiral}), sending participants back into a phase of discovery and improvisation with the altered object. In doing so, the system avoided stagnation and maintained a sense of continual unfolding---an interactive journey that could continue as long as there were new nudges to introduce.
 
\section{Discussion}

  

\subsection{Live Nudging Collective Movements and Shared Agency}

While prior work has shown that MR nudging can bias trajectories through situated virtual overlays~\cite{Kasahara2025MR,Williams2025Nudging,Duives2025Nudging}, our study shifts from the single-person, goal-oriented, quantitative focus of prior MR nudging research toward a collective, open-ended, embodied, qualitative analysis of group movement in co-located settings. 

\subsubsection{MROs as Live Nudges}
Through our study of RQ1, we explored how MROs can nudge collective movement among co‑located participants without prescribing specific actions. We found that \emph{intentional nudging is grounded on neutral affordances}: Affordances describe action possibilities relative to a body in context~\cite{Gibson2015ecological}. By introducing MROs that leverage familiar physical metaphors (rope, spring, magnetic field), we provide \emph{neutral} affordance grounding that easily matches our intuition and quickly supports embodied mastery, while \emph{nudging} adds the designer's (or OJ's) intention by selectively amplifying certain contingencies. For example, a rope that visually and sonically behaves like a heavy cord invites instinctive swinging. OJs then map \emph{``digital physics''} and bodily action to synesthetic audiovisual feedback through the MROs, intentionally making some moves \emph{``rewarded''} more richly than others (e.g., 360° spins for rope), a nudge that biases exploration toward specific actions. For example, the pitch of a tone was mapped to the rope's centripetal acceleration, so that faster, more vigorous swinging produced a higher-pitched sound. MR nudging preserves voluntariness and open-endedness, offering an alternative path between motion guidance based on feedforward/corrective cues toward a target~\cite{Yu2024Design}, and actuation-based puppeteering that reduces agency~\cite{Lopes2022Editorial}.

Our findings also revealed that \emph{synesthetic feedback is highly effective for MR nudging}. Because participants cannot physically touch MROs, synesthetic audiovisual mappings~\cite{Merter2017Synesthetic,Haverkamp2013Synesthetic} function as perceptual proxies for materiality, enabling people to see and to hear \emph{``digital physics''}, tension, momentum, and field strength. In line with research on multimodal feedback in motor learning~\cite{TurmoVidal2023Intercorporeal, Diller2024Visual}, these tight action-effect couplings create virtuous cycles: movement produces clear audiovisual changes that reveal the MRO's capabilities, which then encourage more adventurous movement. These mappings teach the MRO's digital physics without explicit instruction, scaffolding the experience from initial tentative exploration to embodied mastery.

\subsubsection{Nudging the Soma, Not Just the Body or the Mind}
In the introduction, we distinguished two extremes: motion guidance that instructs the mind with corrective cues toward a target movement~\cite{Yu2024Design,Diller2024Visual,Elsayed2022Understanding}, and actuation-based puppeteering that forces the body to comply~\cite{Lopes2022Editorial,Lopes2017Providing,Tamaki2011PossessedHand,Patibanda2024SharedFusion}. The first treats the body as an executor of cognitive decisions; the second bypasses cognition altogether. Both implicitly split mind from body. MR live nudging, by contrast, adopts the non-dualistic stance of somaesthetics~\cite{Shusterman1999Somaesthetics,Hook2018Designing,Hook2021Unpacking,Loke2018Somatic}, operating at the level of the \emph{soma}---the living, sentient body-mind treated as an indivisible unity~\cite{Hook2017SomaBased}. Rather than instructing participants what to do or forcing their limbs to move, the OJ reshapes the felt environment---its digital physics, audiovisual contingencies, and affordance landscape---so that participants naturally perceive, interpret, and respond with their whole embodied selves. The nudge works precisely because it does not separate a ``decision to move'' from ``the movement itself''; instead, perception of the MRO's dynamics and bodily response arise together as a single somatic act, thus realizing the very meaning of \citet{Thaler2008Nudge}'s nudge---an environmental alteration that predictably steers behavior without forbidding options or requiring deliberation~\cite{Thaler2008Nudge,Caraban201923a}---where the body-mind moves along the path of least friction, freely and without coercion. Echoing \citet{Perovich2024Feeling}'s ``Feeling Data through Movement'', participants described ``hearing the force'' and instinctively leaning into it, rather than following an instruction or being pulled: the digital physics invited the soma, and the soma answered.

\subsubsection{Shared Agency}

The mediation of MROs in GravField led to shared authorship and fluid agency and entanglement. The performance resulted in shared control: Participants' movements shaped the actual sonic outcomes; the OJ configured mappings and parameters; and the MROs exhibited quasi-autonomous dynamics that participants treated as responsive \emph{``others.''} This produced \emph{shared authorship}~\cite{Patibanda2024SharedFusion}, where control was neither centralized nor static but \emph{fluid}, shifting with attention and opportunity. Agency~\cite{Tapal2017Sense} was continuously negotiated among participants, OJs, and MROs. This highlighted the fluid power dynamic: participants were not merely being led by the technology or the OJ, but could also influence the unfolding moment-by-moment interaction through their actions, with the OJ ready to adapt and follow their lead. Such negotiated authorship and agency resonate with the entanglement theory of HCI~\cite{Frauenberger2019Entanglement} and soma design's non-dualistic stance~\cite{Hook2021Unpacking,Stepanova2024Intercorporealb}.

\subsubsection{The Emergence of Coordination and Meaning}
Our study extends work on spatial nudging in physical environments~\cite{Grisiute2024Spatial} into co-located MR for multiple participants. We observed that groups would spontaneously fall into synergistic patterns. \emph{Coordination emerged through self-organized synchrony}: the MROs provided situated cues that biased collective movement in certain directions. These coordinated movements sometimes could not be predicted by the OJ, though they occasionally aligned with expected nudging outcomes. For example, dyads converged on mirrored pulses with the rope and oppositional tension, while trios developed relay-like sequences and inclusion strategies. These patterns align with social entrainment and synchrony effects in group coordination~\cite{PhillipsSilver2010Ecology,daSilva2024How,Gordon2025Interpersonal} and with dance improvisation practices of scores and props to structure but not determine movement~\cite{Kimmel2018Sources}. Importantly, these dynamics were not imposed by the system but emerged from local negotiations through the affordances of the MRO and other participants, consistent with enactive accounts of intercorporeal coupling~\cite{Tanaka2015Intercorporeality}.

GravField's live reconfiguration capabilities leverage \emph{ambiguity in MROs as a core design resource}~\cite{Gaver2003Ambiguity, Boon2018Ambiguity, TurmoVidal2024Ambiguity} for open-ended nudging. Rather than steering the group toward a single predefined outcome, the system maintained a state of metastable play~\cite{Tognoli2014Metastable}: patterns of coordinated movement formed, dissolved, and re-formed as new object behaviors were introduced. Unclear or surprising feedback from MR objects compelled participants to interpret and adapt rather than simply execute known actions. When the behavior of the rope MROs changed unexpectedly, participants negotiated and made sense of the new dynamics, often developing novel movement vocabularies. These ambiguities and perturbations prevented premature convergence on any one routine, keeping the group in a poised, exploratory state (a kind of participatory metastability). Through these ongoing adjustments, collective movement self-organized: rather than being externally choreographed, coordination emerged from continuous social coupling and mutual responsiveness within the group. This observation resonates with enactive theories of intersubjectivity~\cite{Fuchs2009Enactive}, which emphasize how coupled intentional actions between people create shared meaning~\cite{Fischer-Lichte2008Transformative}.

\subsubsection{Sustaining the open-endedness}
Through abstraction of soma trajectory across sessions, we found a recurrent arc pattern we termed the \emph{Live Nudging Spiral}. We explain this phenomenon using the concept that \emph{metastability sustains interest at the edge of predictability}~\cite{Werner2007Metastability}. Both participants and OJs seek to maintain continuous novelty to sustain the Live Nudging Spiral: participants develop new movement patterns while OJs reconfigure the system to maintain engagement and excitement. As familiarity increases, interestingness plateaus~\cite{Silvia2005What}, prompting OJ interventions (parameter shifts, remappings) that defamiliarize the field to reopen exploration. This cyclic interplay of familiarity and novelty echoes open-endedness in computational creativity~\cite{Soros2024Creativitya} and prompts from contact improvisation~\cite{Pallant2006Contact}. This phenomenon can be explained by extended neural metastability in an embodied model of sensorimotor coupling as proposed by~\citet{Aguilera2016Extended}: metastability is not restricted to synchronization in neural assemblies but extends to the entire system composed of interactions between brain, body, and environment. As with a DJ, it can be hard to tell whether the OJ is actively intervening or simply playing preset configurations~\cite{Montano2010How}. However, our findings show that pre-recorded MROs alone cannot sustain true open-endedness—live intervention remains essential.

\begin{figure*}[ht]
  \centering
  \includegraphics[width=0.8\linewidth]{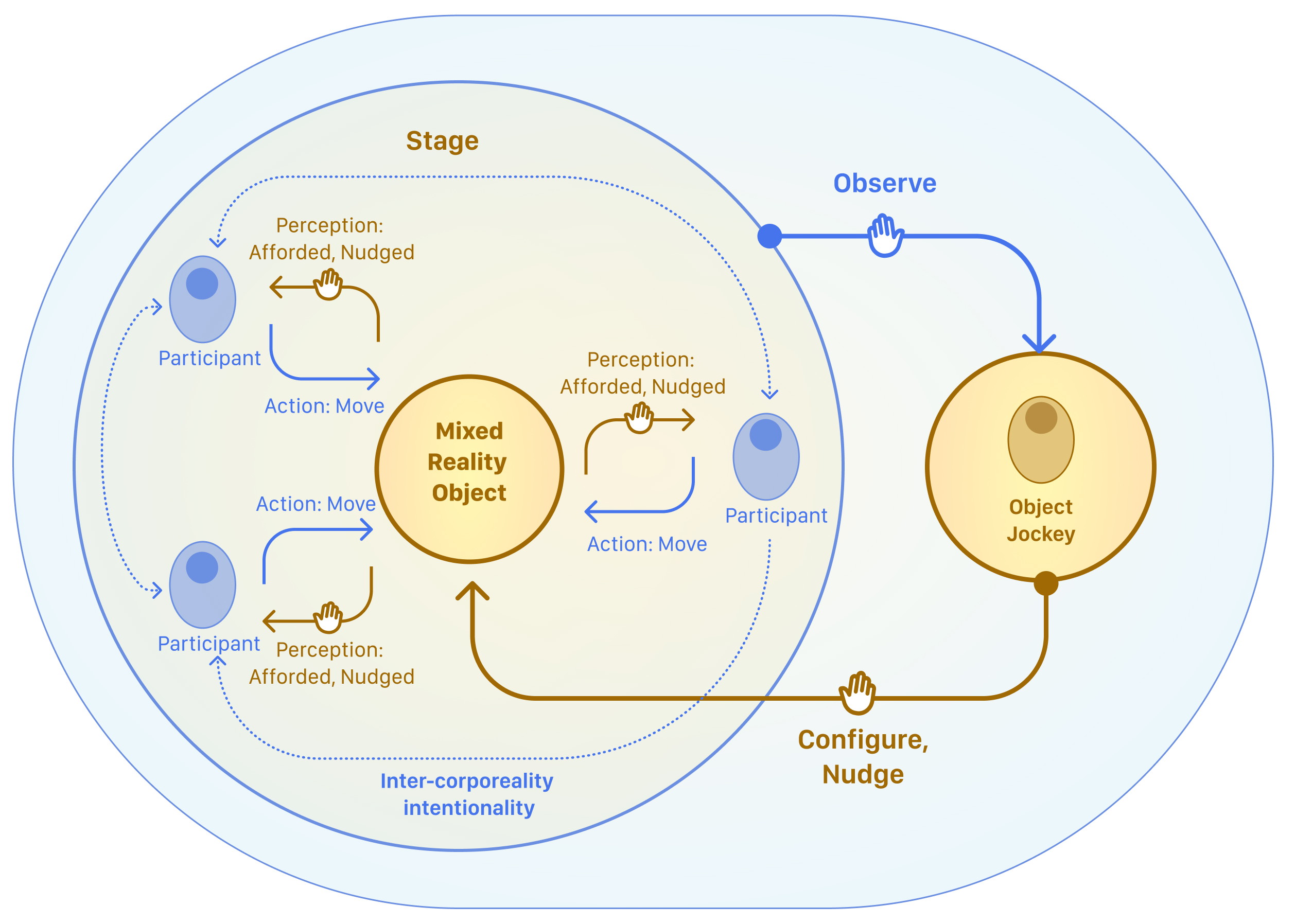}
  \caption{MR-Mediated Intercorporeal Perception--Action Loop. Participants perceive and move through shared MROs, creating intercorporeal coupling, while the Object Jockey observes and live-configures the affordance landscape.}
  \label{fig:assemblages}
  \Description{Conceptual diagram of the MR-Mediated Intercorporeal Perception--Action Loop. A central Mixed Reality Object node connects to multiple Participant nodes arranged in a circle around a shared stage area; each Participant node shows a "perceive" arrow from the object (labeled afforded and nudged) and a "move" arrow back to the object, forming individual loops. Between Participant nodes, bidirectional arrows labeled "intercorporeal intentionality" represent mutual bodily influence. Outside the main loop, an Object Jockey node connects via an "observe" arrow from the stage and a "configure" arrow back to the Mixed Reality Object, representing real-time parameter adjustment.}
\end{figure*}

\subsection{Strategies for Designing MR Live Nudges}

We synthesize these findings into the concept of \textit{Mixed Reality Live Nudges}, grounded in the \emph{MR-mediated intercorporeal perception--action loop} (see Figure~\ref{fig:assemblages}) as an analytical model: live-configured MR objects mediate a shared \emph{``digital physics''} that gently biases collective movement while preserving participant-led sensemaking, even though virtual objects are intangible. Participants act and perceive through MRO and intercorporeal information, while facilitators continuously observe and reconfigure the affordance landscape, enabling designers (or facilitators) to reprogram mixed reality~\cite{Suzuki2025Programmable} on the fly.

However, critics of nudging argue that even non-coercive environmental alterations can undermine autonomy by bypassing reflective deliberation~\cite{DeRidder2024Simple,Wachner2020How}. This concern is amplified when nudges target the soma rather than cognition alone: interfaces that shape bodily actions blur the line between persuasion and coercion. \citet{Benford2012Uncomfortable}'s \emph{``uncomfortable interactions''} further highlight how embodied technologies raise issues of consent and agency. The \emph{programmability} of MR~\cite{Suzuki2025Programmable} magnifies both opportunity and risk: reconfigurable physics can delight, instruct, or manipulate at scale in public or semi-public spaces. We need norms for programmable reality that govern body-influencing interventions in social settings. Participants should be informed about movement influences, maintaining autonomy within the nudge. Designers should adopt \emph{``ethics-by-design''}~\cite{Dignum2018Ethics}, incorporating safeguards for user agency, physical safety, appropriate body contact, consent mechanisms, and accommodations for diverse abilities.

From this, we propose the following design strategies for MR live nudges:

\paragraph{Start with grounded metaphors, then bend them.}
Begin from metaphors aligned with everyday \emph{intuitive physics} (e.g., springs, ropes, magnetism) so newcomers can couple quickly with low entry friction. Once this coupling is established, deliberately \emph{bend} the mapping, tilt the affordance landscape toward preferred actions by introducing non‑linear gains, asymmetric friction, one‑way valves, or hysteresis. This preserves comprehensibility while gently biasing behavior toward the designer’s intended coordinative moves.

\paragraph{Compose synesthetic mappings that “sound out” effort.}
Because MROs are intangible, couple action--effort contingencies to sound to create tight \emph{as‑if haptic} loops. Map intensity to amplitude envelope, direction change to timbre, potential energy to pitch, and proximity to spatialization; add brief transients for contact and release. These synesthetic contingencies support perceptual learning without physical haptics and invite whole‑body engagement to \emph{``feel''} the \emph{``digital physics''}.

\paragraph{Design for metastability and open‑endedness.}
Avoid single‑objective target states. Keep the system poised in a \emph{metastable} regime with multiple shallow ambiguous attractors and gentle perturbations (e.g., novelty injection, rotating micro‑intentions). In our sessions, patterns of coordinated movement formed, dissolved, and re‑formed as new object behaviors were introduced, compelling participants to interpret and adapt rather than execute fixed routines---\emph{participatory metastability}~\cite{Aguilera2016Extended}.

\paragraph{Balance legibility and ambiguity.}
The design goal is to make mapping understandable while also maintaining an element of intrigue. This can be achieved by providing an intuitive visual representation of the body’s interactions, complemented by more ambiguous or exploratory auditory feedback. By balancing these elements, participants can intuitively grasp their bodily interactions while encouraging exploration and interpretation of more complex and evolving auditory stimuli. Calibrated discomfort can be meaningful~\cite{Benford2012Uncomfortable}.

\paragraph{Leverage proxemics as a nudging mechanic.}
Expose intercorporeal relational variables, distance, bearing, and relative height, and make their thresholds and consequences salient so that proxemics becomes a first‑class material for play~\cite{Mueller2014Proxemics}. For example, increase field strength when dyads align heading, open gates when triads reach a distance triangle, or attenuate forces when bodies cluster too tightly. Small, legible changes in relational state should yield noticeable, learnable shifts in MRO behavior.

\paragraph{Ethics‑by‑design for nudging movement.}
When designing MR nudging to guide behavior, designers should consider ethical implications. This includes implementing safeguards for user agency, safety, consent mechanisms, appropriate body contact, and respecting reality's norms. For instance, rope MROs should not swing too quickly, as this could create safety concerns during movement.

\subsection{Limitations and Future Work}
\label{sec:limitation}

Our study highlights several areas where the work could be extended. 

A central limitation is that our OJ role was effectively \emph{``fixed''}: we worked with only two OJs, both of whom were deeply involved in the system's creative development. GravField fuses multiple tools (Unity, Ableton, TouchOSC), and in practice the learning curve makes ad-hoc onboarding difficult within short workshop timelines. This constrains replicability and may reflect the specific sensibilities of these particular OJs rather than the system's general potential. Future work should recruit and train a broader pool of OJs and develop an \emph{``OJ playbook''} (e.g., documented intervention patterns, parameter presets, safety rails, and logging of when/why changes occur) to make the practice more transferable across facilitators and venues.

One challenge lies in how nudging is typically evaluated. Most studies rely on task-based, quantitative metrics, yet in our open-ended setting, such measures risk overlooking the very qualities we sought to investigate. A promising direction would be to combine quantitative traces with participants’ accounts of sensemaking, enabling behavioral shifts to be interpreted alongside how they were experienced.  

Another limitation comes from the sensing scope. By focusing mainly on head movement, we anchored feedback to a single locus of control. This narrowed the embodied repertoire available to participants and restricted the nuance of interaction. Extending the system to whole-body sensing, through depth cameras, inertial sensors, or optical skeleton tracking, could distribute forces across limbs, diversify mappings, and broaden accessibility for different movement abilities.  

The ecology of MROs also remains relatively narrow. We chose three metaphors, rope, spring, and magnetic field, because they provided intuitive starting points. While effective, this limited set constrains generalizability. Future explorations could expand the repertoire of MROs and articulate parameter templates for each, diversifying the range of interactions and making the system more adaptable across cultural contexts, group sizes, and venues.  

Spectator involvement presents another avenue for development. In our workshops, audience members could view the interaction only through static perspectives, which conveyed the visuals but not the felt qualities of participation. Designing dynamic spectator views or lightweight interactive roles could make the system more engaging as a performance medium and extend the social reach of the experience.  

Beyond technical and design aspects, normative questions deserve attention. As mixed reality nudging enters everyday programmable environments, issues of consent, governance, and venue-specific policies will become pressing. Establishing community frameworks and ethical guidelines will be essential to ensure that co-located MR nudging remains safe, inclusive, and socially responsible.

\section{Conclusion}

How can virtual objects shape collective movement without scripting actions or overriding agency? This paper offers one answer through \textit{GravField}, an open-source co-located MR performance system in which an Object Jockey live-configures virtual objects—ropes, springs, magnetic fields—to nudge intercorporeal movement. Through performance workshops with 25 participants, we showed that digital physics grounded in intuitive metaphors, made tangible through synesthetic audiovisual mappings that let participants ``hear'' and ``see'' intangible forces, can invite emergent coordination, sustain productive ambiguity, and support fluid negotiation of agency among participants, facilitators, and MR objects. Our empirical analysis revealed a recurrent \emph{Live Nudging Spiral}—a six-stage cycle of learning, internalizing, coordination, exploration, boredom, and reconfiguration—through which OJs sustain open-ended engagement by defamiliarizing the affordance landscape at the edge of predictability. From these findings, we proposed \emph{MR Live Nudges} as a concept for co-located MR, described the \emph{MR-mediated intercorporeal perception--action loop} as an analytical model of how they operate, and derived six design strategies for non-prescriptive nudging. We see this work as a step toward \emph{programmable realities}~\cite{Suzuki2025Programmable}—where designers compose digital physics, not just interfaces, to foster collective embodied sensemaking.

\begin{acks}
This research was funded by Holo Interactive, Inc. We thank Xiaobo Aaron Liu for developing the GravField system and assisting with the workshop. We thank Raul Masu from Hong Kong University of Science and Technology (Guangzhou) and Ray LC from the School of Creative Media at City University of Hong Kong for their advice on the manuscript. We thank Gottfried Haider, the organizers of the International Conference on Live Coding, and New York University Shanghai for hosting our workshop. We thank Echo Zhou and Esther Zhou from the University of Washington, and Yue Li and Bingqing Chen from Xi'an Jiaotong University for helping organize participants. We thank all participants for their enthusiastic engagement in our workshop at the International Conference on Live Coding.

\textbf{Disclosure of LLM usage.} We used ChatGPT and Claude to support the preparation of this manuscript. This usage included: turning Excel format tables into \LaTeX{} format tables, polishing existing writing, proofreading for grammar and spelling, citation verification, and camera-ready formatting.
\end{acks}
%
\bibliographystyle{ACM-Reference-Format}
\balance
\bibliography{reference}

\clearpage

\appendix

\section{Design Process and System Details}

This appendix provides the full design narrative behind GravField: the embodied practice that inspired the metaphors (\S\ref{sec:ci_inspiration}), the iterative prototyping stages through which the system took shape (\S\ref{sec:design_iterations}), the audiovisual design rationale and formal design principles (\S~\ref{sec:av_rationale}), and detailed specifications for each MRO (\S\ref{sec:mro_principles}) and the OJ intervention model (\S\ref{sec:oj_model}).

\subsection{From Contact Improvisation to MRO design}
\label{sec:ci_inspiration}

The design of GravField's MR objects is deeply rooted in our team's personal experience with Contact Improvisation (CI)~\cite{Paxton1975Contact, Pallant2006Contact}---a dance form in which partners continuously negotiate touch, weight, and shared momentum without predetermined choreography. In the first author's CI class, we were introduced to the practice of using \textit{imagined objects} to mediate bodily interactions: the leading instructor encouraged us to imagine an object, such as a rope, connecting us to our dance partner. This rope could either be loose, allowing for free movement, or taut, pulling us closer together or guiding our motion. The varying tension within this imagined object served as a metaphor for the dynamics of our interaction, influencing the way we responded to each other's movements.

In those classes of CI, the presence of a live musician proved crucial. The musician's improvisation responded to our bodily movements, subtly guiding our actions with changes in tempo, dynamics, and rhythm. At times the musician would alter the music to trigger new movements or behaviors, creating an evolving feedback loop between sound, movement, and interpretation. This experience underscored the power of both physical and abstract objects---whether real or imagined---in shaping embodied interactions, and it directly informed the OJ role in GravField.

As we advanced in CI classes, we incorporated tangible physical objects such as tensioned textiles, which further emphasized the role of materiality in shaping movement. These objects regulated and enhanced bodily interactions, adding layers of complexity and texture. Through these encounters, we recognized the potential of both imagined and physical objects in facilitating collective body movement and shaping social dynamics.

To extend these insights into mixed reality, we sought to replicate the CI experience digitally. We designed MROs that leverage familiar physical metaphors---ropes, springs, and magnetic fields---to guide and influence interactions between participants, just as a rope or tensioned textile in CI creates varying dynamics of connection and separation. The system's MROs dynamically alter the ``tension'' between participants using digital metaphors that stretch or contract based on proximity, movement, and collaboration. William Forsythe's concept of ``\emph{choreographic objects}''~\cite{forsythe2011choreographic}---that objects and spatial constructs can serve as tools encouraging creative exploration of body--space dynamics and intercorporeal possibilities---and specifically his \textit{Lectures from Improvisation Technologies}~\cite{Forsythe1999Improvisation}\footnote{\url{https://improvisation-technologies.zkm.de/}}, a series of video lectures\footnote{\url{https://www.youtube.com/watch?v=Vx0fe9R1D7E}} in which animated lines and geometric overlays reveal how movement can be analyzed as spatially inscriptive, offered further design inspiration. GravField extends this idea into mixed reality, using digital objects as virtual choreographic tools whose properties can be tuned live, enabling participants to experience embodied interaction in a mediated, multi-sensory context.

Through collaborative brainstorming sessions with CI practitioners, we gathered further insights into how physical objects influence movement and interaction. The rope, for example, serves as a mediator for shared balance and tension; the spring introduces ideas of elasticity and resistance; the magnetic field taps into the concept of invisible forces that create attraction or repulsion. These three metaphors were selected because they offer distinct relational dynamics, are immediately legible from everyday physics, and remain open enough for improvisation and live remapping by the OJ.

\subsection{Four-stage iterative development}
\label{sec:design_iterations}

Guided by the CI-informed rationale above, we developed GravField through four prototyping stages, each adding a layer of complexity and yielding reflections that shaped the next iteration.

\paragraph{Stage 1: Single-player, sound-only (embodied sketching for ``digital physics'').}
\textbf{Goal:} establish a synesthetic action--sound loop strong enough that a participant can build an embodied intuition of ``digital physics'' without visuals or haptics.
\textbf{What we built:} a minimal pipeline where one participant's 6DoF head pose streamed to an OSC layer; we mapped simple pose-derived features (height, distance-to-origin, velocity) onto a small set of sound modulators in Ableton Live (e.g., filter cutoff, amplitude, grain density).
\textbf{Key reflection:} random or decorative sounds failed to produce a stable perception--action coupling. The participant could not ``read'' the system. In contrast, metaphor-consistent mappings (e.g., more distance $\rightarrow$ thicker/denser sound; faster motion $\rightarrow$ louder/brighter sound) enabled the participant to anticipate how movement would change sensation, which became the basis of ``digital physics'' as a felt relation. Importantly, we did not write custom simulation code of digital physics in this stage; we treated Ableton as a rapid ``sketching surface'' for somatic mapping.

\begin{figure}[h]
  \centering
  \includegraphics[width=\linewidth]{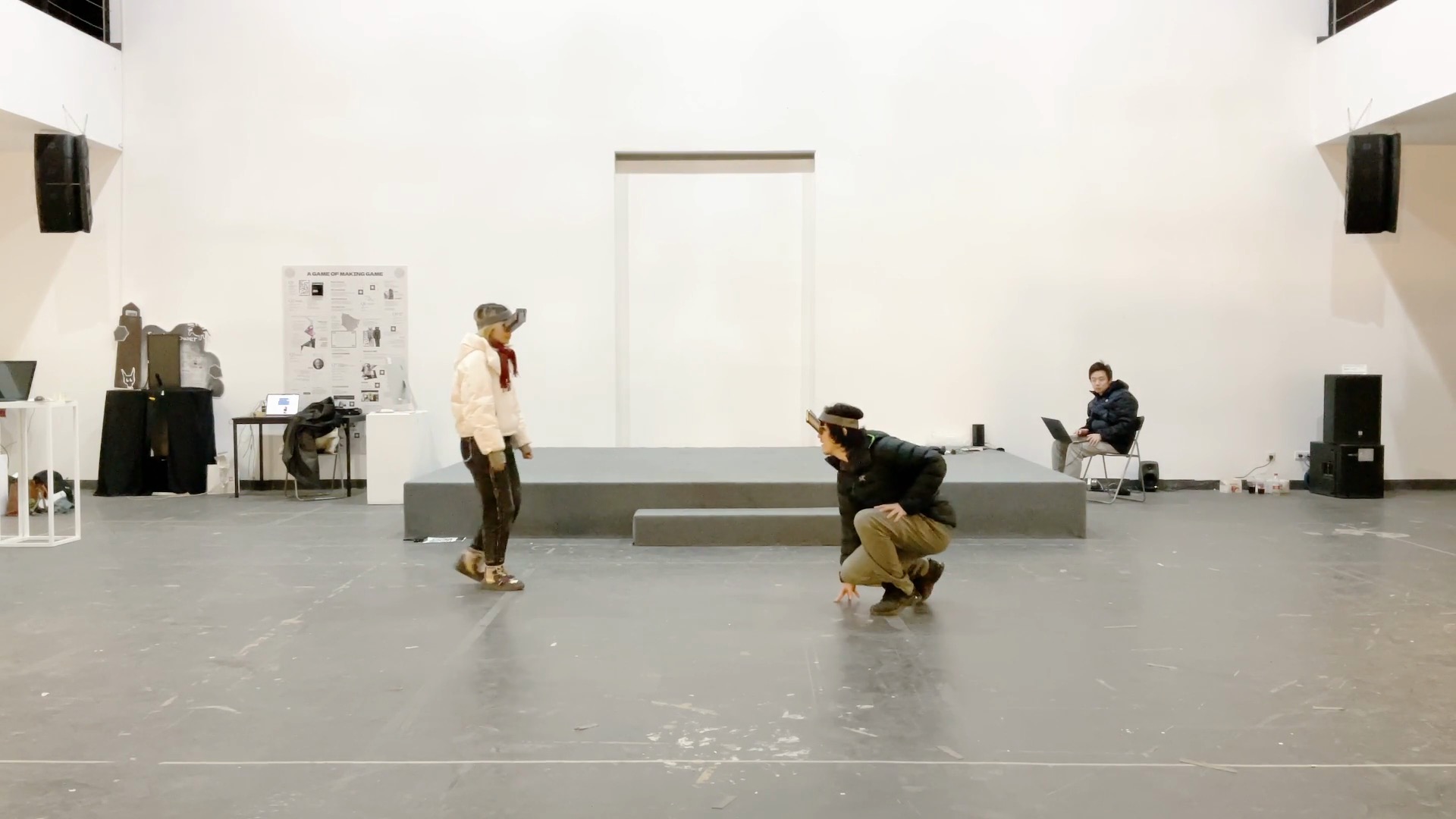}\\[4pt]
  \includegraphics[width=\linewidth]{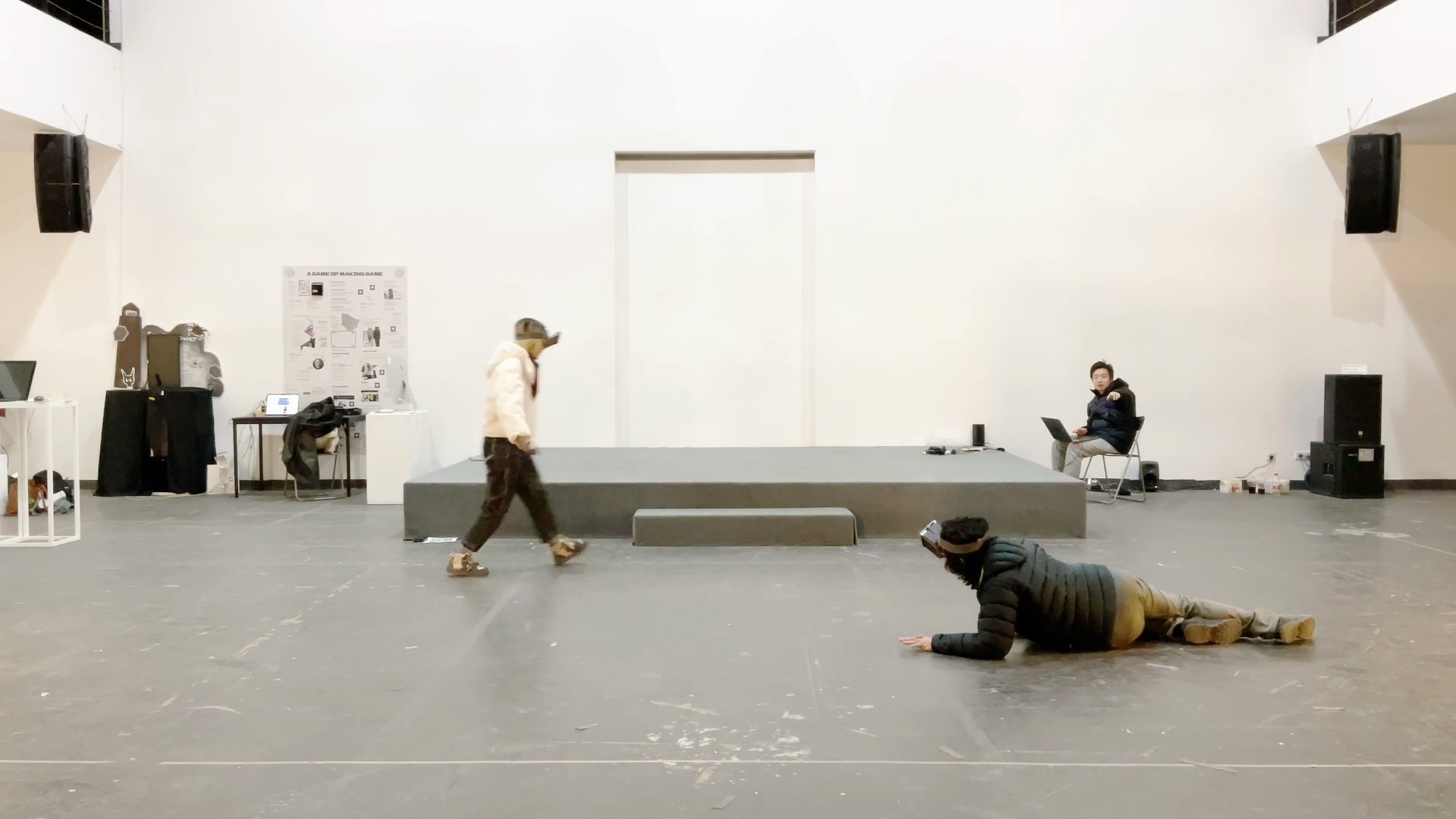}
  \caption{Stage~2 early testing: multi-player, sound-only exploration. Two participants wearing head-mounted devices improvise movement in a shared space while relational audio feedback responds to their spatial configuration.}
  \label{fig:stage2_testing}
  \Description{Two side-by-side photographs from an early Stage 2 testing session. Both show two participants wearing head-mounted displays moving freely in a large studio space, with a facilitator observing from a raised platform in the background. In the left image, one participant stands while the other crouches low. In the right image, one participant walks while the other lies on the floor, demonstrating the range of movement exploration during sound-only multi-player testing.}
\end{figure}

\paragraph{Stage 2: Multi-player, sound-only (discovering the importance of intercorporeal signals).}
\textbf{Goal:} test whether spatial relations between multiple bodies (without visuals) can become a shared material for group sensemaking.
\textbf{What we built:} we added the co-location system, still without MR visuals (see Figure~\ref{fig:stage2_testing}). The orchestration layer computed relational features such as inter-participant distance, relative height, and pose derivatives (speed). These were sent via OSC for sound synthesis/modulation.
\textbf{Key reflection:} intercorporeal and spatial information (e.g., closing distance, height differences, coordinated acceleration) produced emergent social meaning---participants could ``agree'' on what the system was doing by hearing how the collective configuration changed. This stage clarified that GravField is fundamentally a relational system: the most meaningful signals are not individual states, but \textit{between-body} variables.

\paragraph{Stage 3: Multi-player with MR visuals (physical-object metaphors as constraints).}
\textbf{Goal:} add visual form that narrows the interpretive space enough to invite intuitive movement, without prescribing choreography.
\textbf{What we built:} we integrated Unity-based MR rendering and brainstormed $\sim$15 candidate metaphors (e.g., chains, cloth, bubbles, flows, swarms). We discarded options that were too ambiguous to read quickly, too complex to compute robustly from head pose, or too difficult to explain through audio.
\textbf{Key reflection:} without a guiding metaphor, participants explored an effectively infinite movement space and struggled to converge on shared ``rules.'' Physical-object metaphors acted as productive constraints: they anchored interpretation (``what is this like?''), which in turn anchored action (``how should I move with it?''). This became a core rationale for the final MRO set.

\paragraph{Stage 4: Multi-player with MR visuals + live configuration (making nudging performative).}
\textbf{Goal:} support in-the-moment orchestration so the facilitator can sustain open-ended exploration and reconfigure affordances as the group stabilizes.
\textbf{What we built:} we added a TouchOSC-based control surface and a live orchestration workflow where object parameters and mappings could be tuned in real time. We formally treated OSC as a shared ``bus'' between MR simulation, the OJ interface, and the audio engine.
\textbf{Key reflection and convergence:} live configuration changed the nature of the experience: rather than a fixed interactive artifact, GravField became a \textit{performable system}. We converged on three MROs---Rope, Spring, Magnetic Field---because they (i) are quickly legible, (ii) can be computed robustly from head-pose-only inputs, (iii) offer distinct relational dynamics, and (iv) are easily ``played'' by the OJ through parameter changes and audio remapping.

\subsection{Audiovisual design rationale}
\label{sec:av_rationale}

The iterations above confirmed that tightly coupled audiovisual feedback is essential: participants need to \textit{perceive} digital physics in order to act on it. A core challenge is that, unlike physical props, MR objects can be seen but not physically touched. Our participants do not wear haptic gloves or exoskeletons; the only output channel beyond vision that we reliably have is sound. The key question, then, was how to make participants \textit{feel} digital physics---the weight of a rope, the tension of a spring, the pull of a magnetic field---without haptic hardware.

A crucial inspiration came from Ge Wang's repurposing of game controllers as embodied musical interfaces~\cite{wang2009stanford, Huberth2016Notation}, where, for example, the Gametrak's retractable tethers let performers physically pull strings to directly modulate sound. Wang's work showed that metaphor-consistent mappings between bodily action and sonic feedback can create a compelling sense of physicality: the tether feels like a musical string not because it resists like one, but because it \textit{sounds} like one. We adopted this principle for GravField: if sound can make a plastic tether feel like a vibrating string, then a virtual rope rendered in MR, coupled with responsive audio, could make participants feel as though they are swinging a weighted object together.

Building on this insight, we couple synesthetic sounds with augmented visuals to invoke sensations of physical qualities such as weight, tension, or speed. For example, when a virtual spring is stretched, the combination of a rising audio pitch and a stretched visual gives a powerful sense of tension, prompting participants to respond as if a real force were present. Similarly, elastic ropes exhibit a ``wobble'' effect, and heavier ropes appear thicker and droop more. Our aim was for participants to act \emph{as if} they were feeling real forces, closing a perception--action loop through perceptually coherent crossmodal feedback. We also gave the OJ the ability to remap these couplings in real time because the power of nudging lies in being able to subtly alter how objects feel, prompting participants toward new behaviors. For example, linking rope speed to sound amplitude rewards vigorous swings, while mapping spring distance to pitch encourages negotiation of spacing.

\subsection{MRO specifications}
\label{sec:mro_principles}
\label{sec:mro}

Virtual objects lack physical presence, so we implemented MROs as tightly coupled audiovisual entities. Each MRO exposes a small set of digital-physics parameters that simultaneously drive its visual animation and the sonic mappings used in performance. Each object is structured around four design elements:
\begin{itemize}
  \item \textbf{Visual representation:} a visual metaphor that suggests an interaction model immediately;
  \item \textbf{Configurable parameters:} properties the OJ can adjust live via the orchestration interface;
  \item \textbf{Movement-derived variables:} signals computed from participants' head poses (e.g., distance, velocity); and
  \item \textbf{Sound mappings:} default crossmodal couplings that make the digital physics perceptible.
\end{itemize}
Across all MROs, the system uses finite-difference estimates of motion:
\[
\mathbf{v}(t) = \frac{\mathbf{p}(t)-\mathbf{p}(t-\Delta t)}{\Delta t}, \quad
\mathbf{a}(t) = \frac{\mathbf{v}(t)-\mathbf{v}(t-\Delta t)}{\Delta t}
\]
where $\mathbf{p}(t)$ is a participant's head position in the shared coordinate frame.

\subsubsection{Rope MRO}
\label{sec:rope_mro}

The Rope MRO explores embodied collaboration through the phenomenological concept of \textit{motor intentionality}~\cite{Merleau2008World}, wherein bodily movement is purposefully directed toward objects and other bodies in a pre-reflective manner. Within GravField, the Rope functions as a mediating artifact that structures dyadic coordination, prompting participants to engage in rhythmic swinging motions tightly coupled to real-time auditory feedback.

\textbf{Visual representation.} The Rope is rendered as a semi-flexible virtual tether anchored at the chest position of two participants. Its visual geometry sways, bends, and droops dynamically in response to participant movement (Figure~\ref{fig:rope}).

\textbf{Configurable parameters.} The OJ can adjust the rope's \textit{mass} and \textit{maximum width} in real time via the orchestration interface. Increasing mass imparts a heavier, more inertial quality to the rope's motion, while reducing mass yields a lighter, more immediately responsive dynamic.

\textbf{Movement-derived variables.} The system continuously computes the velocity ($v$) and acceleration ($a$) of the rope's midpoint, derived from the tracked head poses of both participants.

\textbf{Sound mappings.} In the default mapping configuration, velocity modulates amplitude (faster swinging motion produces greater volume), while acceleration modulates pitch (rapid changes in velocity raise pitch). This mapping is designed to evoke the kinesthetic experience of swinging a physical rope, wherein auditory intensity and frequency rise with more vigorous collaborative effort.

Throughout the interaction, the OJ retains the capacity to dynamically reconfigure both the physical parameters and the variable-to-sound mappings, ensuring that the auditory environment remains responsive to and supportive of the participants' evolving intercorporeal negotiation.

\subsubsection{Spring MRO}
\label{sec:spring_mro}

The Spring MRO is designed to explore proximity, drawing inspiration from the concept of interpersonal distance and spatial relationships in social psychology. Proximity is not merely about physical closeness but also involves how bodies perceive and relate to space and others within it. As participants move closer or farther apart, the pitch changes, reflecting the tension and compression of a physical spring (Figure~\ref{fig:spring}). Additionally, microphone volume or vocalizations are mapped to spring width parameters, encouraging participants to combine active voice by speaking or clapping.

\textbf{Visual representation.} The Spring is represented by vibrating sine waves connecting the chest positions of two participants, stretching and compressing as they move, rendered with gradient colors.

\textbf{Configurable parameters.} The OJ can adjust \textit{wave width}, \textit{wave offset}, and \textit{maximum wave length}, modifying both visual appearance and physical sensitivity.

\textbf{Movement-derived variables.} The system tracks interpersonal distance ($d$), individual heights ($h$), and height difference between participants.

\textbf{Sound mappings.} Initially, distance controls pitch (increasing distance raises pitch, simulating tension; decreasing distance lowers it), while height modulates the Low-Frequency Oscillator (LFO) rate, affecting sound modulation. When one participant crouches while another stands, the height difference increases LFO rate, creating tremolo effects in the sound; as participants move closer together, pitch drops, evoking an auditory sense of ``compression.'' This real-time auditory adjustment creates a nuanced and engaging shared experience that is continuously shaped by the participants' movements. As with all MROs, the OJ can change the sound mappings in real time during performance.

\subsubsection{Magnetic Field MRO}
\label{sec:magnetic_mro}

The Magnetic Field MRO explores the forces of attraction and repulsion between individuals, mediated through their bodily interactions and spatial awareness. It creates a simulated magnetic field that responds to the proximity and relative orientation of participants, facilitating the sensation of forces acting between them. This object creates a spatial magnetic vector field that simulates the push--pull dynamics between participants (Figure~\ref{fig:magnetic}).

\textbf{Visual representation.} The Magnetic Field MRO creates an ambient particle field around all participants. Red particles represent positive poles; cyan particles represent negative poles. Particles flow, cluster, or disperse based on magnetic relationships, with chest positions as pole centers.

\textbf{Configurable parameters.} The OJ can assign each participant a magnetic pole (positive or negative) and adjust field strength. Flipping polarity transforms attraction into repulsion. For example, the OJ can suddenly flip one participant's polarity, transforming convergence into repulsion and shifting the sonic texture.

\textbf{Movement-derived variables.} The system computes pairwise distances ($d$) between all participants and measures field turbulence intensity. These parameters control the glitch intensity and texture of the background music.

\textbf{Sound mappings.} Initially, distance and field turbulence control the ``glitchness'' of the background music: closer proximity and greater turbulence produce more fragmented, distorted sound textures. When three participants gradually converge in a circle, particles concentrate at the center and sound becomes fragmented and dense; as they disperse, particles diffuse outward and sound normalizes. When participants are closer together and their magnetic poles align or oppose each other, the sound becomes more ``glitchy,'' reflecting the increased tension and dynamic nature of their relationship within the magnetic field. As the distance increases or orientations become more neutral, the sound normalizes, symbolizing a reduction in tension.

With more than two participants, a three-body problem emerges---creating chaotic, unpredictable particle behavior that establishes an ambiguous space for exploration. By adjusting the relative strengths or flipping polarities, the OJ can gently encourage converging, dispersing, or rotating formations.

\subsection{OJ intervention model}
\label{sec:oj_model}

Conceptually, the Object Jockey (OJ) plays a role similar to a DJ in a club set. Rather than selecting tracks and filters, the OJ selects MROs and manipulates their digital physics to shape the energy, rhythm, and structure of the session. OJ intervention operates at two layers:
\begin{itemize}
  \item \textbf{Physics layer (OJ1):} change object parameters (e.g., make the Rope heavier, the Spring stiffer, the Field stronger) to perturb participants' current embodied coupling and reopen exploration.
  \item \textbf{Mapping layer (OJ2):} alter how movement-derived variables are sonified (e.g., re-route a tension signal from pitch to rhythm) to shift what participants attend to and how ``forces'' are perceived.
\end{itemize}
If no OJ intervenes, GravField can still run with fixed presets; however, live orchestration enables responsive pacing and smoother transitions, sustaining open-ended exploration as group dynamics evolve.

Across all three MROs, mappings are designed to be legible but not fully determined: participants can quickly grasp basic cause--effect relationships while retaining room for exploration. The OJ may adjust both mapping parameters and relations in real time during performance.
 
\section{Semi-Structured Interview Questions}
\label{sec:questions}

\subsection{Post-session  Interview (15--20 min, immediately after MR round)}
\textbf{Setup:} Show a short video replay (30--90s) of the group’s performance. Invite participants to create a soma-trajectory sketch on a single timeline.

\paragraph{Soma-trajectory sketch prompt}
\begin{itemize}
  \item Draw a timeline of your experience from start to finish.
  \item Mark changes over time in: \emph{interestingness/engagement}, \emph{familiarity/understanding}, \emph{sense of agency}, and \emph{connection to others}.
  \item Circle “turning points” (moments when something changed) and annotate what happened.
\end{itemize}

\paragraph{Perceiving “digital physics”}
\begin{itemize}
  \item What did the object feel like (e.g., heavy/light, tense/loose, sticky/elastic)? What cues gave you that impression?
  \item Did sound help you sense forces or dynamics? Any specific mapping you noticed (e.g., pitch, volume, timbre)?
  \item Did you ever act \emph{as if} a force were real (pulling, resisting, yielding)? Can you point to a moment?
\end{itemize}

\paragraph{Intercorporeal coordination and negotiation}
\begin{itemize}
  \item How did you coordinate with others—explicitly (body signal/gestures) or implicitly (timing/entrainment)?
  \item Did roles emerge (leader/follower)? Did that shift?
  \item Did you feel your movement choices were influenced by others’ proximity, height, speed, or attention?
\end{itemize}

\paragraph{OJ nudging and agency}
\begin{itemize}
  \item Did you notice moments when ``digital physics'' changed? What changed, and how did you respond?
  \item Did those changes feel like suggestions, constraints, rewards, or disruptions?
  \item At any point, did you feel you were negotiating with the OJ?
\end{itemize}

\paragraph{Boredom, plateau and re-ignition}
\begin{itemize}
  \item Was there a moment when exploration plateaued (repetition, fatigue, “I’ve tried everything”)? What were the cues?
  \item What helped you re-engage (OJ changes, a partner move, a new interpretation, audience energy)?
\end{itemize}

\paragraph{Witnessing and being watched}
\begin{itemize}
  \item Did knowing others were watching affect how you moved?
  \item Did the spectator's view influence how you understood what was happening?
\end{itemize}

\paragraph{Closing}
\begin{itemize}
  \item If you compare this to your prior creative practice, what felt new or different?
  \item What would you change about the object(s), mappings, or OJ interventions?
\end{itemize}

\subsection{Final Group Discussion (45--60 min)}
\begin{itemize}
  \item Compare experiences across the three MROs: what kinds of coordination did each invite?
  \item How did the presence/absence of clear metaphor shape exploration?
  \item How did OJ interventions affect agency and group dynamics? When were they helpful vs disruptive?
\end{itemize}


\end{document}